%% file: main.tex
\documentclass[a4paper,11pt]{article}
\usepackage{jheppub} 
\usepackage{lineno}
\usepackage{booktabs}
\usepackage{textcomp}

\newcommand{\D}[1]{{\rm d}}
\newcommand{\Br}{{\mathcal B}}

\title{\boldmath Workshop summary -- Kaons@CERN 2023}

\usepackage{comment}

\author[1,2]{G.~Anzivino,}
\author[3]{S.~Arguedas Cuendis,}
\author[4]{V.~Bernard,}
\author[5]{J.~Bijnens,}
\author[6]{B.~Bloch-Devaux,}
\author[7]{M.~Bordone,}
\author[2,8]{F.~Brizioli,}
\author[9]{J.~Brod,}
\author[10,11]{J.M.~Camalich,}
\author[8]{A.~Ceccucci,}
\author[2]{P.~Cenci,}
\author[12]{N.H.~Christ,}
\author[13]{G.~Colangelo,}
\author[14]{C.~Cornella,}
\author[15]{A.~Crivellin,}
\author[16]{G.~D'Ambrosio,}
\author[17]{F.~F.~Deppisch,}
\author[18]{A.~Dery,}
\author[19]{F.~Dettori,}
\author[7]{M.~Di~Carlo}
\author[20]{B.~D\"obrich,}
\author[21]{J.~Engelfried,}
\author[22]{R.~Fantechi,}
\author[23]{M.~Gonz\'alez~Alsonso,}
\author[24]{M.~Gorbahn,}
\author[25]{E.~Goudzovski,}
\author[18]{Y.~Grossman,}
\author[26]{N.~Hermansson-Truedsson,}
\author[27]{Z.~Hives,}
\author[13]{M.~Hoferichter,}
\author[13]{B.-L.~Hoid,}
\author[25,27]{T.~Husek,}
\author[15]{G.~Isidori,}
\author[7,28,29]{A.~J\"uttner,}
\author[27]{K.~Kampf,}
\author[22]{S.~Kholodenko,}
\author[30]{M.~Knecht,}
\author[27]{M.~Koles\'ar,}
\author[27]{M.~Koval,}
\author[25]{C.~Lazzeroni,}
\author[31]{Z.~Ligeti,}
\author[7,32,33]{F.~Mahmoudi,}
\author[34]{R.~Marchevski,}
\author[35,36]{D.~Mart\'inez Santos,}
\author[8,37]{K.~Massri,}
\author[8]{T.~Momb\"acher,}
\author[38]{H.~Nanjo,}
\author[32]{S.~Neshatpour,}
\author[39,40]{T.~Nomura,}
\author[23,41,42,43]{E.~Passemar,}
\author[44]{L.~Peruzzo,}
\author[2]{M.~Piccini,}
\author[23]{A.~Pich,}
\author[28]{C.T.~Sachrajda,}
\author[45]{S.~Schacht,}
\author[39,40]{K.~Shiomi,}
\author[7]{P.~Stangl,}
\author[15,46]{P.~Stoffer,}
\author[8,47]{J.~Swallow,}
\author[7]{J.~T.~Tsang,}
\author[48]{G.~Valencia,}
\author[44]{R.~Wanke,}
\author[9,49,50]{J.~Zupan}

\affiliation[1]{Dipartimento di Fisica e Geologia dell'Universit\`a di Perugia, Via A. Pascoli - 06123 Perugia, Italy}
\affiliation[2]{INFN Sezione di Perugia, Via A. Pascoli - 06123 Perugia, Italy}
\affiliation[3]{Universidad Estatal a Distancia, San Jose, Costa Rica}
\affiliation[4]{IJCLab, Univ.\ Paris-Saclay, CNRS/IN2P3, 91405 Orsay, France}
\affiliation[5]{Division of particle and nuclear physics, Department of Physics, Lund University, Box 118, SE 221 00 Lund, Sweden}
\affiliation[6]{Dipartimento di Fisica dell'Universit\`a di Torino, 10125 Torino, Italy}
\affiliation[7]{Theoretical Physics Department, CERN, Geneva, Switzerland}
\affiliation[8]{CERN, European Organization for Nuclear Research, CH-1211 Geneva 23, Switzerland}
\affiliation[9]{Department of Physics, University of Cincinnati, Cincinnati, OH 45221, USA}
\affiliation[10]{Instituto de Astrof\'isica de Canarias,  C/ V\'ia L\'actea, s/n E38205 - La Laguna, Tenerife, Spain}
\affiliation[11]{Universidad de La Laguna, Departamento de Astrof\'isica, La Laguna, Tenerife, Spain}
\affiliation[12]{Department of Physics, Columbia University, New York, NY 10027, USA}
\affiliation[13]{Albert Einstein Center for Fundamental Physics, Institute for Theoretical Physics, University of Bern, Sidlerstrasse 5, 3012 Bern, Switzerland}
\affiliation[14]{PRISMA$^+$ Cluster of Excellence {\em \&} MITP, Johannes Gutenberg University, Mainz, Germany}
\affiliation[15]{Physik-Institut, Universit\"at Z\"urich, Winterthurerstrasse 190, CH–8057 Z\"urich, Switzerland}
\affiliation[16]{INFN Sezione di Napoli, Complesso Universitario di Monte S. Angelo ed. 6 via Cintia, 80126, Napoli, Italy}
\affiliation[17]{Department of Physics and Astronomy, University College London, London WC1E 6BT, United Kingdom}
\affiliation[18]{Department of Physics, LEPP, Cornell University, Ithaca, NY 14853, USA}
\affiliation[19]{Università degli Studi di Cagliari and INFN Sezione di Cagliari, Cagliari, Italy }
\affiliation[20]{Max-Planck-Institut f\"ur Physik, Boltzmannstrassee 8, 85748 Garching bei M\"unchen, Germany}
\affiliation[21]{Instituto de F\'{\i}sica, Universidad Aut\'onoma de San Luis Potos\'{\i}, 78000, Mexico}
\affiliation[22]{INFN Sezione di Pisa, Largo B. Pontecorvo 3 - 56127 Pisa, Italy}
\affiliation[23]{Department of Theoretical Physics, IFIC, University of Valencia -- CSIC, E-46071 Valencia, Spain}
\affiliation[24]{Department of Mathematical Sciences, University of Liverpool, Liverpool, L69 7ZL, UK}
\affiliation[25]{School of Physics and Astronomy, University of Birmingham, Edgbaston, Birmingham, B15 2TT, UK}
\affiliation[26]{The Higgs Centre for Theoretical Physics, School of Physics and Astronomy, The University of Edinburgh Mayfield Rd, Edinburgh, EH9 3JZ, UK}
\affiliation[27]{Institute of Particle and Nuclear Physics, Charles University, Prague, Czech Republic}
\affiliation[28]{School of Physics and Astronomy, University of Southampton, Southampton, SO17 1BJ, UK}
\affiliation[29]{STAG Research Centre, University of Southampton, Southampton, SO17 1BJ, UK}
\affiliation[30]{Centre de Physique Th\'{e}orique, CNRS/Aix-Marseille Univ./Univ. de Toulon (UMR7332), CNRS-Luminy Case 907, 13288 Marseille Cedex 9, France}
\affiliation[31]{Lawrence Berkeley National Laboratory, University of California, Berkeley, CA 94720, USA}
\affiliation[32]{Universit\'e de Lyon, Universit\'e Claude Bernard Lyon 1, CNRS/IN2P3, Institut de Physique des 2 Infinis de Lyon, UMR 5822, F-69622, Villeurbanne, France}
\affiliation[33]{Institut universitaire de France (IUF), 75005 Paris, France}
\affiliation[34]{École Polytechnique Fédérale de Lausanne (EPFL), CH-1015 Lausanne, Switzerland}
\affiliation[35]{Axencia Galega de Innovaci\'on, Conseller\'ia de Econom\'ia e Industria, Xunta de Galicia, Santiago de Compostela, A Coru\~na, Spain}
\affiliation[36]{Instituto Galego de Fisica de Altas Enerxias, 15705 Santiago de Compostela, A Coruña, Spain}
\affiliation[37]{Physics Department, Lancaster University, Lancaster, LA1 4YW, UK}
\affiliation[38]{Department of Physics, Osaka University, Toyonaka, Osaka 560-0043, Japan}
\affiliation[39]{Institute of Particle and Nuclear Studies, High Energy Accelerator Research Organization (KEK), Tsukuba, Ibaraki 305-0801, Japan}
\affiliation[40]{J-PARC Center, Tokai, Ibaraki 319-1195, Japan}
\affiliation[41]{Department of Physics, Indiana University, Bloomington, Indiana 47405, USA}
\affiliation[42]{Center for Exploration of Energy and Matter, Indiana University, Bloomington, Indiana 47408, USA}
\affiliation[43]{Theory Center, Thomas Jefferson National Accelerator Facility, Newport News, Virginia 23606, USA}
\affiliation[44]{Johannes Gutenberg Universit\"at Mainz, D-55099 Mainz, Germany}
\affiliation[45]{Department of Physics and Astronomy, University of Manchester, Manchester M13 9PL, United Kingdom}
\affiliation[46]{Paul Scherrer Institut, 5232 Villigen PSI, Switzerland}
\affiliation[47]{INFN Laboratori Nazionali di Frascati, I-00044 Frascati, Italy}
\affiliation[48]{School of Physics and Astronomy, Monash University, Wellington Road, Clayton, VIC-3800, Australia}
\affiliation[49]{Berkeley Center for Theoretical Physics, University of California, Berkeley, CA 94720, USA}
\affiliation[50]{Theoretical Physics Group, Lawrence Berkeley National Laboratory, Berkeley, CA 94720, USA}

\preprint{CERN-TH-2023-206}
\abstract{Kaon physics is at a turning point -- while the rare-kaon experiments NA62 and KOTO are in full swing, the end of their lifetime is approaching and the future experimental landscape needs to be defined. With HIKE, KOTO-II and LHCb-Phase-II on the table and under scrutiny, it is a very good moment in time to take stock and contemplate about the opportunities these experiments and theoretical developments provide for particle physics in the coming decade and beyond. This paper provides a compact summary of talks and discussions from the Kaons@CERN 2023 workshop, held in September 2023 at CERN.}

\begin{document}
\maketitle
\flushbottom

\section{Introduction}
The NA62 experiment at CERN and KOTO at J-PARC Japan are the only two experiments worldwide fully dedicated to the study of rare kaon decays. NA62 is planned to conclude its efforts in 2025, and both experiments are aiming to meet important milestones on that time scale. 
The future experimental landscape for kaon physics beyond this date has not taken shape yet, but there is a strong and engaged community committed to continuing these investigations in the coming years. Proposals for next-generation experimental facilities HIKE~\cite{HIKE-Proposal} at CERN and KOTO-II~\cite{Aoki:2021cqa} at J-PARC are on the table and under scrutiny.
With this background, the aim of this workshop was to bring together theoretical and experimental kaon physicists to reflect on the present situation, future challenges and the main goals of the community. 

Kaons, the mesons containing one strange and either a lighter up or down quark, have historically played a central role in developing and establishing the Standard Model (SM) of elementary particle physics. Many of the SM’s salient features were discovered through the study of kaons. For example, parity violation was hinted at in kaon decays~\cite{Lee:1956qn}, kaons were central to the development of the Cabibbo theory of flavour~\cite{Cabibbo:1963yz}, and the absence of flavour-changing neutral currents (FCNCs) at tree level led to the postulation of a fourth (the charm) quark~\cite{Glashow:1970gm}. CP violation, one of the three necessary ingredients to justify the baryon asymmetry of the Universe, was discovered in its direct and indirect incarnations in kaon decays~\cite{Christenson:1964fg,KTeV:1999kad,NA48:1999szy}. It was incorporated in the ``new'' SM by Kobayashi and Maskawa~\cite{Kobayashi:1973fv} by introducing a third generation of quarks before its experimental discovery~\cite{E288:1977xhf}. The Cabibbo--Kobayashi--Maskawa (CKM) quark-mixing matrix describes all quark decays and  is the subject of a major particle physics experimental programme.

The full particle content of the SM was later experimentally established at CERN with the Higgs discovery~\cite{CMS:2012qbp,ATLAS:2012yve} in 2012. Since then, the outlook for particle physics has changed considerably. While the observed baryon asymmetry, the question about the origin of neutrino masses and the patterns of quark and lepton masses and mixings, or the presence of dark matter in the universe, are still lacking a microscopic and confirmed understanding within or beyond the SM (BSM), clear indications of the direction of journey, like hitherto the Higgs particle, are also currently lacking. 

Kaon physics plays a very special role in this context. The study of rare kaon decays provides a unique sensitivity to New Physics (NP), 
than reached by
 collider experiments. In the SM, the rare decay of a charged or neutral kaon into a pion plus a pair of charged or neutral leptons is hugely suppressed. This is due to the absence of tree-level FCNC interactions (e.g., $s\to d$) in the SM. Such a transition can only proceed at loop level involving the creation of at least one very heavy (virtual) electroweak (EW) gauge boson. Two ingredients lead to a massive suppression of the decay rate: the Glashow--Iliopoulos--Maiani (GIM) mechanism, which leads to a suppression of the transition by the heavy-mass scale of the gauge bosons, and the smallness of the involved combination of CKM-matrix elements. Both make rare kaon decays even more suppressed than the rare $B$-meson decays currently studied at LHCb and Belle-II.

While this suppression constitutes a formidable experimental challenge in identifying the decay products amongst a variety of background signals, NP, with mass scales much heavier than the EW scale, could leave a significantly measurable imprint through tree-level or loop contributions. Despite these challenges, nature has been kind to us: rare kaon decays are one of the theoretically cleanest places to search for the effects of NP – one could even say that they constitute a standard candle of the SM. This is on the one hand due to the limited number of possible decay channels of kaons and pions and, as a result, the relatively clean experimental environment. More importantly, and very much in contrast to rare $B$-meson decays, there are “gold-plated” rare decay modes amongst the rare-kaon decays, which are purely short-distance dominated and therefore allow for very precise theory predictions. These are the rare decays of charged and neutral kaons into pions and a pair of neutrinos, $K^+\to \pi^+\nu\bar \nu$ and $K_L\to \pi^0\nu\bar\nu$.

The charged-kaon decay is currently being studied at the NA62 experiment at CERN, and a measurement of its branching ratio with a precision of 15\% is expected by 2025.
However, to substantially improve this measurement, thereby substantially increasing the likelihood of a discovery, the experimental precision will need to be reduced further to the level of the theory prediction, i.e., 5\%. This can only be achieved with a next-generation experiment. The HIKE experiment, a future high-intensity kaon factory at CERN currently under approval, will reach the 5\% precision goal on the measurement of $K^+$ during its first phase of operation. Afterwards, a second phase with a neutral $K_L$ beam aiming at the first observation of the very rare decays $K_L\to \pi^0\ell^+\ell^-$ is foreseen. KOTO-II, a planned but not yet funded evolution of KOTO, 
aims to measure the branching ratio of $K_L\to\pi^0\nu\bar\nu$ with a precision of 25\%.  

With the setup and detectors optimised for the measurement of the most challenging rare-decay processes, HIKE phase 1 and 2 as well as KOTO-II will be able to reach unprecedented precision on many other $K^+$ and $K_L$ decays as well, many of which are also extremely interesting in view of the possibility to provide a window on NP contributions. 
The LHCb experiment will also contribute to kaon physics, especially with studies of $K_S$ decays.
What makes them now less appealing than the golden modes, is the fact that long-distance effects are more relevant or in some cases even dominating, so that NP effects may be hidden behind poorly understood hadronic effects. But significant progress is happening on that front too. For instance, the communities working on lattice QCD, effective field theory and dispersive approaches have, over the last decades, continued sharpening their tools motivated in part by the wealth of experimental information on kaon decays, which calls for a deep and precise theoretical understanding of the hadronic contributions. The prospects for further improving our control over non-perturbative effects on the same time scale as the planned new experiments are very good, as presentations and discussions during the workshop have made clear.

This workshop summary aims to present a concise overview of the current status of experimental and theoretical kaon physics, to discuss opportunities and expectations for future developments and improvements in precision and to provide entry points into the vast literature on the subject, and is structured as follows. Section~\ref{sec:exp} is dedicated to experimental aspects and provides an overview of current experiments and of planned future ones. Section~\ref{sec:SM} summarises the current situation and prospects for improvement of our understanding of kaon decays within the SM. It briefly touches upon the remarkably broad spectrum of quantum field theory tools which have been developed and have to be used in connection with kaon decays. Section~\ref{sec:BSM} is dedicated to a discussion of the huge potential of rare kaon decays for the discovery of NP, in light of the current situation and the future prospects of indirect searches at $B$ factories and direct searches at the energy frontier. The complementarity with these searches provides a strong motivation for carrying out this programme. In Sec.~\ref{sec:conclusions}, general conclusions and an outlook are provided.
\include{experimental}

\section{Kaon physics in the Standard Model}
\label{sec:SM}
Rare kaon decays proceed through 
FCNCs that are suppressed in the SM~\cite{Glashow:1970gm}. They thus offer unique
possibilities to discover indirect evidence of degrees of freedom that describe physics beyond the SM, and therefore remain a very active and exciting field of research for both theory and experiment. The neutral- and charged-kaon decay modes $K\to\pi\nu{\bar\nu}$ 
stand out among this class of processes, being entirely dominated by the contributions from short-distance scales. This
allows for very precise SM predictions as detailed in Sec.~\ref{sec:gold-plated theory}. The situation also looks promising for the class
of radiative kaon decays, like $K\to\gamma^{(*)}\gamma^{(*)}$, $K\to\pi\gamma^{(*)}$, $K\to\pi\gamma^{(*)}\gamma^{(*)}$, \ldots,
where the photon(s) can be either real or virtual, and in which case long-distance hadronic effects represent an important contribution. To predict them in the SM one uses simulations of lattice QCD~(Sec.~\ref{sec:Lattice}), ChPT and 
dispersion theory (Secs.~\ref{sec:ChPT} and~\ref{sec:ChiPT and dispersion}). A kaon factory naturally also produces a flux of pions and the option of studies of its decay channels is addressed in Sec.~\ref{sec:pi0}.
Strategies for finding observables, e.g., $K_S$--$K_L$ interferences in the time dependence of the decay probability,
that are free from the uncertainties due to hadronic contributions can be devised (cf.\ Sec.~\ref{sec:Grossman}). Their 
experimental implementation remains challenging.
Section~\ref{sec:SM_predictions} summarises some of the reflections made in the discussion session on kaons in the SM,
including the interplay between lattice QCD, Sec.~\ref{sec:Lattice}, ChPT, dispersion relations, and short-distance constraints, Secs.~\ref{sec:ChPT} and~\ref{sec:ChiPT and dispersion}, using as example promising new insights into resolving the long-distance contributions in the rare $K_L\to\ell^+\ell^-$ decay.

\subsection{Theory calculations for the gold-plated modes in the SM}\label{sec:gold-plated theory}
The rare kaon decays $K^+ \to \pi^+ \nu \bar{\nu}$ and $K_L \to \pi^0 \nu \bar{\nu}$ are among the cleanest probes of physics beyond the SM.
They are generated by highly virtual EW box and $Z$-penguin diagrams that can be calculated to high precision in perturbation theory.
The leading decay matrix elements can be extracted from precisely measured semileptonic kaon decays, while GIM suppressed light-quark contributions are tiny and calculable both in ChPT and on the lattice.
The leading contribution to the two rare $K \to \pi \nu \bar{\nu}$ decays is captured by the effective Hamiltonian~\cite{Buchalla:1995vs}
\begin{align}
\label{eq:HeffSM}
  \mathcal{H}_{\text{eff}} = \frac{4G_F}{\sqrt{2}}
  \frac{\alpha}{2\pi\sin^2\theta_w} \sum_{\ell=e,\mu ,\tau}
  \left( \lambda_c X^\ell + \lambda_t X_t \right)
  (\bar{s}_L \gamma_{\mu} d_L)
  ( \bar{\nu}_{\ell L} \gamma^{\mu} \nu_{\ell L}) + \text{h.c.}\,,
\end{align}
where the dependence on the relevant CKM matrix elements appears as the coefficient $\lambda_i = V_{is}^* V_{id}^{\phantom{8}}$ in front of the short-distance Wilson coefficients $X^\ell$ and $X_t$.

The top-quark contribution $X_t$ is a function of $x_t = m_t(\mu_t)^2/M_W^2$ and has been calculated including two-loop QCD~\cite{Buchalla:1998ba,Misiak:1999yg} and EW~\cite{Brod:2010hi} corrections.
All perturbative corrections to $X_t$ only involve the scale $\mu_t$, where $\alpha_s$ is small.
This suggests an excellent convergence of the perturbation series, which is confirmed by the preliminary results of the three-loop calculation \cite{GSY-xt:preparation}.

The analytical expressions for $X_t$, as well as the numerical value of $m_t$ and $M_W$, depend on the QCD and EW renormalisation schemes.
The $\overline{\text{MS}}$ scheme is the natural choice regarding QCD. The numerical value $m_t(m_t) = 162.83(67)\,$GeV is obtained from the top-quark pole mass (see Ref.~\cite{Brod:2021hsj} for further details).
A numerical value for $X_t$ is obtained by calculating a mean value of the QCD contribution, $X_t^{\text{QCD}}$, by varying $\mu_t \in [60,320]$GeV and adding the EW corrections.
In total, one finds at NLO in QCD and two-loop EW
\begin{equation}
  X_t
  = 1.462 \pm 0.017_{\text{QCD}} \pm 0.002_{\text{EW}}
  \,.
\end{equation}
The theory uncertainty associated with the QCD corrections is given by
the difference of the central value
and the minimal / maximal value in the $\mu_t$ interval.
This uncertainty is expected to reduce even further at next-to-next-to-leading order (NNLO) in QCD.

The charm-quark contribution $X^{\ell}$ is a function of the neutrino flavour $\ell$,
while the parameter $P_c = \lambda^{-4} (\tfrac{2}{3}X^e+\tfrac{1}{3}X^{\tau})$ comprises the charm-quark contribution to $K^+ \to \pi^+ \nu \bar{\nu}$, which involves a sum over all neutrino flavours.
$P_c$ has been calculated at
NNLO in QCD~\cite{Buras:2006gb} and at
NLO in the EW interactions~\cite{Brod:2008ss}. The hard GIM mechanism ensures that it is $x_c = m_c(\mu_c)^2/M_W^2$ suppressed and its numerical value has recently been updated \cite{Brod:2021hsj} to $P_c = (0.2255/\lambda)^4 \times (0.3604 \pm 0.0087)$.
The computation of $P_c$ involves double insertions of charged-current operators that are matched onto the operator of Eq.~(\ref{eq:HeffSM}).
Current conservation ensures that perturbative corrections are absent below the charm scale at this order of the expansion.
The effects of dimension-eight operators at the charm threshold, as well as additional long-distance contributions arising from up- and charm-quarks, have been estimated in Ref.~\cite{Isidori:2005xm}, leading to the correction $\delta P_{c,u} = 0.04(2)$. These effects can be computed using lattice QCD in the future, as discussed in Sec.~\ref{sec:lattice-kpinunu}.

The branching ratio of the charged mode is then given by
\begin{equation}\label{eq:BR:ch}
  \Br \left(K^+\to\pi^+\nu\bar{\nu}(\gamma)\right) = \kappa_+
  (1+\Delta_{\text{EM}})
  \Bigg[\left(\frac{\text{Im}\lambda_t}{\lambda^5} X_t\right)^2 +
  \left(\frac{\text{Re}\lambda_c}{\lambda} \left(P_c + \delta P_{c,u}
    \right) + \frac{\text{Re}\lambda_t}{\lambda^5} X_t\right)^2
  \Bigg]\,,
\end{equation}
where the hadronic matrix element is contained in the parameter $\kappa_+$.
It is extracted from $K_{\ell 3}$ decay including higher-order chiral
corrections~\cite{Mescia:2007kn,Bijnens:2007xa}. The NLO QED
corrections~\cite{Mescia:2007kn} are parameterised by
$\Delta_{\text{EM}} = -0.003$ in Eq.~\eqref{eq:BR:ch}.

The remaining parametric input is contained in the CKM factors
$\lambda_t$ and $\lambda_c$, defined above. In the numerical evaluation, these parameters are expanded
in $\lambda$, including the quadratic corrections \cite{Brod:2021hsj}. The leading order expansion  $\text{Im}\lambda_t = A^2\bar{\eta}\lambda^5$, $\text{Re}\lambda_t = A^2\lambda^5(\bar{\rho} - 1)$ and $\text{Re}\lambda_c = - \lambda$ already involves all Wolfenstein parameters. The PDG \cite{ParticleDataGroup:2022pth} quotes two different sets of numerical values for these parameters, which are based on the methods of the CKMfitter \cite{Hocker:2001xe} and UTfit \cite{UTfit:2005ras} collaboration, respectively. They read
\begin{equation}
  \label{eq:wolf-CKM-UT}
  \lambda =
  \begin{cases}
    0.22499(67) &\\
    0.22500(67) &
  \end{cases}
  A =
  \begin{cases}
    0.833(11) &\\
    0.826^{+0.018}_{-0.015} &
  \end{cases}
  \bar{\rho} =
  \begin{cases}
    0.159(10) & \\
    0.159(10) & \\    
  \end{cases}
  \bar{\eta} =
  \begin{cases}
    0.348(9)  & \text{UTfit} \\
    0.348(10) & \text{CKMfitter}
  \end{cases}
\end{equation}
For these CKM parameters the following prediction for the charged mode in the SM are obtained:
\begin{equation}
  \Br(K^+ \to \pi^+ \nu \bar \nu) =
  \begin{cases}
    8.38(17)(25)(40) \times 10^{-11} & \text{UTfit input} \\
    8.19(17)(25)(53) \times 10^{-11} & \text{CKMfitter input}
  \end{cases} \,.
\end{equation}
The errors in parentheses correspond to the remaining short-distance,
long-distance, and parametric uncertainties, with all contributions
added in quadrature.
In more detail, one finds for UTfit CKM input the leading contributions to the uncertainty as
\begin{equation}
\begin{split}
  10^{11} \times & \Br(K^+ \to \pi^+ \nu \bar \nu)
   = 8.38
          \pm 0.14_{X_t^{\text{QCD}}}
          \pm 0.01_{X_t^{\text{EW}}}
          \pm 0.11_{P_c}
          \pm 0.25_{\delta P_{cu}} \\
 & \qquad
          \pm 0.04_{\kappa_+}
          \pm 0.14_{\lambda}
          \pm 0.31_{A}
          \pm 0.12_{\bar\rho}
          \pm 0.03_{\bar\eta}
          \pm 0.05_{m_t}
          \pm 0.15_{m_c}
          \pm 0.06_{\alpha_s}
\,,
\end{split}
\end{equation}
where the combined error is 6\%. With  huge efforts under way on reducing the dominant residual parametric uncertainties, we expect this error to reduce further over the coming years.
Theory uncertainties are already  smaller and future theoretical calculations will considerably reduce both the short- and long-distance uncertainties. 
%
Here we note the excellent idea to form a ratio of the charged decay mode and $\varepsilon_K$ \cite{Buras:2021nns,Buras:2022wpw} that cancels large parts of the parametric uncertainties. 
This ratio is also theoretically very clean, given recent progress~\cite{Brod:2019rzc,Brod:2021qvc,Brod:2022har,GJS-BK:preparation} in the theory prediction of $\varepsilon_K$, 
{and leads to a precision of about 5\% for both the charged and neutral kaon decay with consistent central values. The cancellation of CKM uncertainties, in particular for the ratio involving the neutral kaon decay, indicates that $\Delta F=2$ processes have a crucial impact on the determination of CKM parameters in global fits.}

The branching ratio of the neutral mode is computed from
\begin{equation}
  \Br\left(K_L\to\pi^0\nu\bar\nu\right)=
  \kappa_L r_{\epsilon_K}
  \left(\frac{\text{Im}\lambda_t}{\lambda^5}X_t \right)^2\,,
  \label{eq:brkL}
\end{equation}
and it depends to a good approximation only on the top-quark function $X_t$
discussed above. The hadronic matrix element is 
again extracted from $K_{\ell 3}$ decay including higher-order chiral
corrections~\cite{Mescia:2007kn}, while $r_{\epsilon_K}$ parameterises the small impact of indirect CP violation~\cite{Buchalla:1996fp}.
$X_t$ and all remaining parametric input have been discussed above in the context of the charged mode. The SM prediction for the neutral mode then reads
\begin{equation}
  \Br(K_L \to \pi^0 \nu \bar \nu) =
  \begin{cases}
2.87(7)(2)(23) \times 10^{-11} & \text{UTfit input} \\
2.78(6)(2)(29) \times 10^{-11} & \text{CKMfitter input}
  \end{cases}
 \,.
\end{equation}
Again, the errors in parentheses correspond to the remaining
short-distance, long-distance, and parametric uncertainties, with all
contributions added in quadrature. In more detail, the leading
contributions to the uncertainty for UTfit CKM parameters are
\begin{equation}
\begin{split}
  10^{11} \times \Br(K_L \to \pi^0 \nu \bar \nu)
 & = 2.87 \pm 0.07_{X_t^{\text{QCD}}}
          \pm 0.01_{X_t^{\text{EW}}}
          \pm 0.02_{\kappa_L}\\
 & \quad  \pm 0.15_{\bar\eta}
          \pm 0.15_{A}
          \pm 0.07_{\lambda}
          \pm 0.03_{m_t}
\,.
\end{split}
\end{equation}

\subsection{Lattice QCD for non-perturbative contributions in  kaon decays}\label{sec:Lattice}
As experimental uncertainties decrease, improving the precision of theoretical calculations becomes crucial to perform reliable tests of the SM. 
Lattice QCD provides a powerful means for non-perturbative, first-principle determinations of several hadronic observables through extensive Monte Carlo simulations. 

\subsubsection{Theory and general methodology}
As the community plans the next generation of studies of rare kaon decays, it may be interesting to reflect on the long timescales required, not only to perform the experiments but also to develop the theoretical methods and carry out the computations. In the history of kaon physics this is nicely illustrated by $K\to\pi\pi$ decays, processes in which both indirect and direct CP violation were first discovered. It is only within the last decade that quantitative results for the amplitudes in lattice computations have been obtained~\cite{Blum:2015ywa,RBC:2015gro,Bai:2016ocm,RBC:2020kdj}, and in particular for the $\Delta I=1/2$ rule (after more than half a century) and the direct CP-violation parameter $\epsilon^\prime/\epsilon$ (after more than two decades). 
The latest lattice QCD result for the $\Delta I=1/2$ rule is  $\mathrm{Re}\,A_0/\mathrm{Re}\,A_2=19.9(2.3)(4.4)$,
where the values in parentheses give the statistical and systematic errors respectively, to be compared to the experimental value of $\mathrm{Re}\,A_0/\mathrm{Re}\,A_2=22.45(6)$.
As we now understand thanks to the lattice results, the surprisingly large value results from a variety of QCD effects including a suppression of $\mathrm{Re}\,A_2$ as well as an enhancement of $\mathrm{Re}\,A_0$\,
\cite{RBC:2020kdj}.
Here $A_0$ and $A_2$ are the decay amplitudes for decays into two-pions with total isospin 0 and 2, respectively. 
For $\epsilon^\prime/\epsilon$ the lattice result is $\mathrm{Re}(\epsilon^\prime/\epsilon)=0.00217(26)(62)(50)_\mathrm{IB}$\,\cite{RBC:2020kdj} to be compared to the experimental value of 0.00166(23)\,
\cite{NA48:2002tmj,KTeV:2010sng,ParticleDataGroup:2022pth}. For the lattice result the first and second error are again statistical and systematic, respectively, and the error with the subscript {\footnotesize IB} corresponds to the uncertainty due to isospin-breaking effects which are amplified because of the $\Delta I=1/2$ rule, for which the central value of the result obtained using ChPT was taken\,\cite{Cirigliano:2019cpi}. The emphasis now is on reducing the computational and theoretical uncertainties.
{
We note that the authors of Ref.~\cite{Bardeen:1986vz}, using the dual QCD approach had also found an enhancement of $\mathrm{Re}\,A_0$ and the suppression of $\mathrm{Re}\,A_2$ (see also Ref.~\cite{Buras:2014maa} for an updated analysis).
The dual QCD approach gives a value for $\epsilon'/\epsilon$ of $0.0005(2)$~\cite{Buras:2022cyc, Buras:2023qaf}, below 
both the experimental result of $0.00166(23)$ and the above lattice result of $0.00217(26)(62)(50)$~\cite{RBC:2020kdj}. A recently updated estimate, 
based on analytical techniques for both short- as well as long-distance effects, gives $0.0014(5)$~\cite{Cirigliano:2019ani}.
}

In the last decade or so the range of physical quantities and processes for which the non-perturbative hadronic effects can be computed using lattice QCD has been extended very significantly. This includes the possibility of evaluating matrix elements of bi-local operators of the form
\begin{equation}\label{eq:bilocal}
\int d^4x \langle f|O_1(x)\,O_2(0)|K\rangle,
\end{equation}
where $O_{1,2}$ are weak or electromagnetic local operators. Applications include the $K_L$--$K_S$ mass difference $\Delta M_K\equiv M_{K_L}-M_{K_S}$ \cite{Christ:2012se,Bai:2014cva,Bai:2018mdv}, the long-distance contributions to the indirect CP-violating parameter $\epsilon_K$~\cite{Christ:2015phf,Bai:2023lkr}, the rare kaon decays $K\to\pi\ell^+\ell^-$, where $\ell=e$ or $\mu$ \cite{Isidori:2005tv,Christ:2015aha,Christ:2016mmq,RBC:2022ddw}, and the long-distance contribution to the golden mode $K^+\to\pi^+\nu\bar{\nu}$
\cite{Christ:2016eae,Bai:2017fkh,Bai:2018hqu,Christ:2019dxu}. 
By ``long-distance" we mean a separation between the operators greater than the inverse charm-quark mass and the lattice computations are therefore performed in four-flavour QCD. This allows us to exploit the GIM mechanism where appropriate to reduce, or even avoid, the additional ultraviolet divergences which potentially arise when $x\to 0$. In addition, the precision of the perturbative matching calculation relating the matrix element of operators renormalised non-perturbatively to the Wilson coefficients calculated in the $\overline{\mathrm{MS}}$ scheme is improved at larger momentum scales, in this case above $m_c$.

For $\Delta M_K=3.483(6)\,$MeV and $\epsilon_K=2.228(11)\times 10^{-3}$~\cite{ParticleDataGroup:2022pth}, for which $O_1$ and $O_2$ are the $\Delta S=1$ effective weak Hamiltonians, it is likely not possible for lattice QCD computations to reach the experimental precision in the next decade. Nevertheless errors of ${\mathcal O}(5\%)$, or perhaps smaller, can be achieved on $\Delta M_K$ and the long-distance contribution to $\epsilon_K$ with adequate computing resources (and in the latter case with improved determinations of $V_{cb}$). At this level of precision, a comparison of the theoretical and experimental results for these very small FCNC quantities will provide significant tests of the SM and constraints on its extensions.

\subsubsection[$K\to\pi\nu\bar\nu$ decays]{$\boldsymbol{K\to\pi\nu\bar\nu}$ decays}
\label{sec:lattice-kpinunu}
While these decays are short-distance dominated, lattice QCD computations can provide a first principles determination of the long-distance effects in $K^+\to\pi^+\nu\bar\nu$ decays with controlled errors. 
{The contribution of these effects to the branching ratio is expected to be $\mathcal{O}(5\%)$ }
(they are negligible for $K_L\to\pi^0\nu\bar\nu$ decays). For these decays, one of the operators in Eq.~\eqref{eq:bilocal} is a $\Delta S=1$ weak operator from the effective Hamiltonian and the other is a $\Delta S=0$ weak operator corresponding to the emission of either a virtual $W$-boson or a virtual $Z$-boson. The theoretical framework has been developed \cite{Christ:2016eae} and been implemented in a number of exploratory numerical studies with unphysical quark masses~\cite{Bai:2017fkh,Bai:2018hqu,Christ:2019dxu}. In the latest study~\cite{Christ:2019dxu} it was found that the momentum dependence of the amplitude was very mild, so that it may be sufficient to compute the amplitude at a limited number of kinematic points, and that the contribution from the two-pion intermediate state can be evaluated but is small (less than 1\%).  The next step is a computation on a $64^3\times 128$ lattice with near-physical meson masses ($M_\pi=135.9(3)\,$MeV and $M_K=496.9(7)$\,MeV) with a target uncertainty on the long-distance contribution of 30\%. Further reductions in the error, towards one of ${\mathcal O}(10\%)$ or less, will require computations at several lattice spacings and are achievable in the next 5--10 years.

\subsubsection[$K_L\to\pi^0\mu^+\mu^-$ decays]{$\boldsymbol{K_L\to\pi^0\mu^+\mu^-}$ decays}

Measuring this partially CP-violating decay is a target of the second phase of HIKE and making a SM prediction for this process will be possible with current lattice QCD methods in the same time frame.  
As explained, for example, in Ref.~\cite{Cirigliano:2011ny} there are three contributions to this decay of approximately equal size:  (i) the rare second-order-weak short-distance process which is the target of these studies, (ii) indirect CP violation proportional to $\epsilon_K$ and the amplitude for the $K_S\to\pi^0\mu^+\mu^-$ decay 
and (iii) the CP-conserving process in which the final state $\mu^+\mu^-$ pair is produced by two photons.  
An accurate result for contribution (ii) will be an automatic outcome of the lattice QCD calculations 
of the amplitude for $K_S\to\pi^0\ell^+\ell^-$ decays, for which the framework enabling lattice computations of the parameters $a_S$ and $b_S$ has been developed\,\cite{Christ:2015aha} and it is expected that they will be evaluated with uncertainties below the 10\% level within the next 5--10 years. In particular the sign of $a_S$ will be determined.
While not yet thoroughly studied, contribution (iii) should also be calculable in lattice QCD, using the methods being developed for the $K_{L,S}\to\mu^+\mu^-$ 
decays discussed in Sec.~\ref{sec:Kls-mumu}, with 10\% accuracy -- a plausible objective in 5--10 years.

\subsubsection[$K^+\to\pi^+\ell^+\ell^-$ decays]{$\boldsymbol{K^+\to\pi^+\ell^+\ell^-}$ decays}

The framework for lattice calculations of the amplitude for $K^+\to\pi^+\ell^+\ell^-$ decays ($\ell=e,\mu$) has been developed~\cite{Christ:2015aha} and exploratory numerical studies have been 
performed~\cite{Christ:2016mmq,RBC:2022ddw}. 
For these decays, one of the operators in Eq.~\eqref{eq:bilocal} is the $\Delta S=1$ weak Hamiltonian and the other is an electromagnetic current.
The emphasis will now be on a reduction of both the statistical and systematic errors with the expectation that the parameters $a_+$ and $b_+$ will be determined with uncertainties below the 10\% level within the next 5--10 years.

\subsubsection[$K_{L,S}\to\ell^+\ell^-$ decays]{$\boldsymbol{K_{L,S}\to\ell^+\ell^-}$ decays}
\label{sec:Kls-mumu}

The framework for the computation of the contribution from the two-photon intermediate state to the complex amplitudes of $K_L\to\mu^+\mu^-$ and $K_S\to\mu^+\mu^-$ decays is being developed~\cite{Christ:2020bzb}. An important first step was a full computation of the complex amplitude of the related decay $\pi^0\to e^+e^-$~\cite{Christ:2022rho}. This calculation was performed on 5 gauge ensembles with inverse lattice spacing ranging from 1.015\,GeV to 2.36\,GeV, so that the continuum limit can be taken, resulting in a result with a precision better that 10\% for both the real and imaginary parts of the amplitude. A second step has been the exploratory numerical study of the decay $K_L\to\gamma\gamma$ with the aim of controlling the subtraction of unphysical exponentially growing terms in the time separation between the weak Hamiltonian and the emission of the first photon~\cite{Zhao:2022pbs}. The focus now is on the calculation of $K_L\to\mu^+\mu^-$ and a first result for the quark-line connected part of this process was presented in Ref.~\cite{ChaoPos2023}.   A result with 10\% accuracy is expected within 5 years. The presence of the $\pi\pi\gamma$ intermediate state introduces a systematic error believed to be below 5\% that will require new 3-body methods to remove.

\subsubsection{Lattice QCD+QED}
Over the past decade, the precision of lattice calculations has advanced to a point where previously neglected subleading effects now demand careful consideration~\cite{FlavourLatticeAveragingGroupFLAG:2021npn}.
These effects include the corrections due to electromagnetic interactions and those related to the up and down quark mass difference, which are both expected to be of~$\mathcal{O}(1\%)$. 
The inclusion of such isospin-breaking effects in lattice simulations is conceptually and computationally challenging, mainly due to the difficulty of defining QED in a finite volume with periodic boundary conditions. Many prescriptions have been formulated over the years~\cite{Hayakawa:2008an,Endres:2015gda,Kronfeld:1990qu,Lucini:2015hfa,Feng:2018qpx,Christ:2023lcc} and applied to the calculation of many different observables. 
The currently observed $3\sigma$ tensions with unitarity in the first row of the CKM matrix~\cite{FlavourLatticeAveragingGroupFLAG:2021npn,ParticleDataGroup:2022pth,Cirigliano:2022yyo} motivated lattice calculations of isospin-breaking corrections to leptonic decay rates of light mesons, with the aim of determining $|V_{us}|$ and the ratio $|V_{us}/V_{ud}|$ with sub-percent precision. 
A theoretical framework for lattice QCD+QED calculations of leptonic decay rates has been first developed by the RM123+Soton collaboration in Ref.~\cite{Carrasco:2015xwa}. This has been successfully applied to the decay rates of pions and kaons into muon-neutrino pairs by the RM123+Soton group~\cite{Giusti:2017dwk,DiCarlo:2019thl} and more recently by the RBC/UKQCD collaboration~\cite{Boyle:2022lsi}, providing results in agreement with each other and with previous ChPT calculations~\cite{Cirigliano:2011tm}. The results of Ref.~\cite{Boyle:2022lsi} highlighted the relevant role of finite-volume effects in this kind of calculations, which scale only as inverse powers of the lattice size and can be potentially sizeable. Work is in progress to tame such systematic uncertainty~\cite{DiCarloPos2023,DiCarlo:2023rlz}, with the goal of reaching a precision on $|V_{us}/V_{ud}|$ below half percent in the next couple of years. This sets a milestone in precision calculations on the lattice and further progress is expected in the near future. A third lattice calculation is in fact currently ongoing, following an alternative method recently proposed in~Ref.~\cite{Christ:2023lcc}. This method differs in the treatment of long-distance QED corrections to the decay amplitudes: as a consequence finite-volume effects are expected to be exponentially suppressed, rather than power-like. 

Applications of lattice QCD+QED are not limited to kaon leptonic decays though, but extensions of this framework to processes with hadrons in the final state are currently under study. This new frontier of calculations includes kaon semileptonic decays,  $K\to \pi \ell\nu$, which can provide an independent estimate of~$|V_{us}|$, and hadronic kaon decays like $K\to\pi\pi$, which is crucial for the study of CP violation in the SM. In both cases a new issue arises, which is related to the analytic continuation from Euclidean to Minkowski space-time of those correlation functions where a photon is exchanged between two particles in the final state. A first theoretical study of QED corrections to $K\to \pi \ell\nu$ on the lattice has been done in Ref.~\cite{Sachrajda:2019uhh}, and more recently in Ref.~\cite{Christ:2023lcc}. Given the current interest in the topic and the impressive recent progress in the field, first lattice results could appear in the next few years. The inclusion of QED corrections in $K\to\pi\pi$, and hence $\mathrm{Re}(\epsilon'/\epsilon)$, is even more complicated because of the possible mixing of the final-state $\pi\pi$ channels with total isospin~0 and~2. Initial studies on this have been done in Refs.~\cite{Cai:2018why,Christ:2017pze,Christ:2021guf},  marking the start of a challenging research avenue that will extend over the next decade.

\subsection[ChPT, short-distance constraints, and large $N_c$]{ChPT, short-distance constraints, and large $\boldsymbol{N_c}$}\label{sec:ChPT}
Besides \textit{ab initio} calculations provided by numerical
simulations of QCD on a discretised space-time (cf.\
Sec.~\ref{sec:Lattice}),
ChPT remains a fundamental tool to study kaon decays in
general~\cite{Cirigliano:2011ny}. In this section, we focus on
radiative kaon decays such as $K\to\gamma^{(*)}\gamma^{(*)}$,
$K\to\pi\gamma^{(*)}$, $K\to\pi\gamma^{(*)}\gamma^{(*)}$, \ldots,
where the photon(s) can be either real or virtual.
In the latter case it materialises as a lepton--antilepton pair,
$\gamma^*\to \ell^+\ell^-$,
$\ell = e, \mu$. The presence of a real or virtual photon generates a
contribution to the decay amplitude from long-distance physics,
i.e., from QCD in the non-perturbative regime, which dominates over
the short-distance part where NP could potentially hide.

This class of radiative decays displays specific features that makes
the interplay with large-$N_c$ arguments and short-distance
constraints particularly important. That is, due to electromagnetic
gauge invariance, the lowest-order contribution to the amplitude often
starts only at NLO in the low-energy
expansion~\cite{Ecker:1987qi,Ecker:1987hd}.
Moreover, the full structure of the amplitude and/or its dependence
with respect to the kinematic variables is then only
revealed at NNLO. Making predictions for
these processes thus requires a theoretical
understanding and a numerical evaluation of the corresponding
LECs. Unfortunately, this knowledge is
lacking at present and this makes predictions difficult.

In the strong sector of ChPT the LECs can be estimated by resonance 
saturation \cite{Ecker:1988te,Cirigliano:2006hb,Bijnens:2014lea}, which finds its justification in the 't Hooft large-$N_c$ limit of QCD, $N_c\to\infty$, 
$\alpha_s N_c \to {\rm const.}$ \cite{tHooft:1973alw}.
In the weak sector, the limit $N_c\to\infty$ has been proposed quite some time ago \cite{Buras:1985yx}
for non-leptonic decays. But a systematic 
study of the large-$N_c$ limit applied to the case of rare kaon decays has not been undertaken yet. It is important to stress
here that the large-$N_c$ limit is not being considered in order to provide an adequate description of the 
whole amplitude. It would certainly fail to do so, since chiral loops, whose contribution to the 
decay distribution is clearly visible in the experimental decay distribution in, for instance,
$K^\pm \to \pi^\pm e^+ e^-$, are suppressed in this limit.
Rather, it is only meant to be used to get a possible handle on the values of the LECs that contribute to a given amplitude \cite{Pich:1995qp,Knecht:1998sp,Bijnens:2000im,Cirigliano:2019cpi}.

In this respect, there are two main differences between the situation in the strong sector 
and the one in the weak sector. First, in the former, 
the large-$N_c$ limit can  
be applied to three-flavour QCD, 
immediately before integrating out
also the light quarks and reaching an effective theory where the only surviving degrees of freedom 
are the light pseudo-scalar states. In the weak sector, the last step before reaching this low-energy
description is also provided by three-flavour QCD, but now supplemented by a set of four-fermion 
operators $Q_I$ modulated by coupling constants $C_I$, usually called Wilson coefficients. These 
additional pieces are the low-energy manifestation of the SM degrees of freedom that
populate the spectrum from the EW scale down to the hadronic scale around 1 GeV, 
where only the three lightest quarks remain as active degrees of freedom. The only input that is required from the SM at 
this $\sim 1$ GeV scale is therefore the list of four-fermion operators $Q_I$ and the values 
of the corresponding couplings $C_I$ at this same scale. The second difference is due to the existence, 
in the weak sector and in the particular case of radiative kaon decays, of short-distance singularities 
that do not show up in the strong sector. These short-distance 
singularities arise in QCD correlators involving the time-ordered product of the electromagnetic current with the 
four-fermion operators $Q_I$, which are relevant for radiative kaon decays. This time-ordered product is singular at short distances \cite{Isidori:2005tv,DAmbrosio:2018ytt}, and it 
is mandatory to understand and correctly address this feature before attempting a determination of the LECs.

This second aspect shows up quite clearly in the large-$N_c$ limit, for instance in the $K\to\gamma^*\gamma^*$
transition form factor. It manifests itself in the form of a contribution involving a vacuum-polarisation
function, which is divergent in QCD. This divergence actually disappears when the two photons 
are real, so that there is no problem in defining the amplitude for $K\to\gamma\gamma$ in the usual way,
in terms of a kaon-to-two-photons transition form factor. But it is present 
as soon as at least one of the photons is off-shell, i.e., in the amplitudes for $K\to\gamma\ell^+\ell^-$
or for $K\to\ell^+_1 \ell^-_1 \ell^+_2 \ell^-_2$. In these cases this divergence is taken care of by a local 
contact contribution provided by the operator $Q_{7V}$, the product of the quark current $({\bar s} d)_{V-A}$ 
with the leptonic vector current, the divergence being absorbed by the renormalisation of the corresponding 
coupling $C_{7V}$. The renormalised vacuum-polarisation function has an asymptotic behaviour proportional 
to $\log(Q^2/\nu^2)$. This differs markedly from the strong sector, where the correlators involved
behave asymptotically as inverse powers of $Q^2$. As a consequence, the usual picture of saturation by 
a single resonance---or even a finite number of resonances---does not work, and one needs to consider 
an infinite tower of narrow resonances \cite{DAmbrosio:2019xph}. However, while the addition of the local contribution from $Q_{7V}$
allows one to define a finite transition form factor for a neutral kaon into two virtual photons, its insertion
into the loop integral that leads to the amplitude for the $K\to\ell^+\ell^-$ decay is now divergent.
Actually, if CP is conserved, this divergence only appears in the amplitude for $K_L\to\ell^+\ell^-$,
and, once minimally subtracted, is of the form 
$\sim  (\log\nu) C_{7V} (\nu) \alpha G_F ({\bar s} \gamma^\mu (1-\gamma_5) d) ({\bar\ell}\gamma_\mu \ell)$.
This structure is reminiscent of a two-loop short-distance contribution, also of order ${\cal O}(\alpha^2 G_F)$,
discussed in Ref.~\cite{Eeg:1996pr}.

The large-$N_c$ limit also offers some interesting insight into the amplitudes for the CP-conserving
processes $K\to\pi\gamma^*\to\pi\ell^+\ell^-$~\cite{Knecht:WIP}. 
In the charged-kaon channel, the amplitude in the large-$N_c$ limit is
dominated by the contributions from the current--current operators $Q_1$ and $Q_2$, whose Wilson coefficients
are of similar size, and much larger than those of the QCD penguin operators, but with opposite signs.
The LEC $a_+$ thus results, in the large-$N_c$ limit, from a large cancellation between 
these two contributions, making a stable prediction difficult without some knowledge of $1/N_c$ suppressed corrections.
Since $Q_2$ does not contribute to the form factor for $K_S \to \pi^0 \gamma^*$ in the large-$N_c$ limit,
this cancellation does not occur and one obtains an unambiguous answer for the LEC $a_S$.
Although its value can only be determined with a relative uncertainty of about $1/N_c \sim 30\%$, its 
sign is fixed without ambiguity, and corresponds to a positive interference between the direct and indirect CP-violating 
components of the amplitude for $K_L\to\pi^0\ell^+\ell^-$, which is rather good news in view of the possibility
to measure this interference in the future, see Sec.~\ref{sec:exp}.

A systematic investigation of all radiative kaon decay modes from the perspective of the large-$N_c$ 
limit of QCD remains to be done, but is under way. Although it may not lead to predictions in
all possible cases, it may nevertheless provide a useful
guide to implementing phenomenological approaches that take some known properties from QCD,
in particular at short distances, into better account.


\subsection{ChPT and dispersion relations}\label{sec:ChiPT and dispersion} 

\begin{table}[t]
	\centering
	\footnotesize
	\begin{tabular}{lcccc}
		\toprule
		 & $10^3 \times L_1^r$ & $10^3 \times L_2^r$ & $10^3 \times L_3^r$ & $\chi^2$/dof \\
		 \midrule
		Dispersive treatment, NLO matching~\cite{Colangelo:2015kha} & $0.51(6)$ & $0.89(9)$ & $-2.82(12)$ & $141/116 = 1.2$ \\
		Dispersive treatment, NNLO matching~\cite{Colangelo:2015kha} & $0.69(18)$ & $0.63(13)$ & $-2.63(46)$ & $122/122 = 1.0$ \\
		 \midrule
		BE14 global fit~\cite{Bijnens:2014lea} & $0.53(6)$ & $0.81(4)$ & $-3.07(20)$ \\
		\bottomrule
	\end{tabular}
	\caption{Results for the LECs at $\mu = 770$~MeV, obtained from matching a dispersive representation of $K_{\ell4}$ form factors to ChPT.}
	\label{tab:LECsFromKl4}
\end{table}

Besides large-$N_c$ considerations and matching to short-distance
constraints, the purview of ChPT can also be extended in combination
with dispersion relations, which allow one to implement the
constraints from analyticity, unitarity, and crossing symmetry, and
thereby unitarise the chiral expansion. This section is focused on
$K_{\ell 4}$ decays ($K \to \pi\pi \ell\nu_\ell$), a prominent example
in which the dispersive evaluation of $\pi\pi$ rescattering
corrections is particularly important.

Leptonic and semileptonic kaon decays play a crucial role in the determination of CKM matrix elements. On the one hand, the ratio of $K_{\ell2(\gamma)}$ (i.e., $K \to \ell \nu_\ell (\gamma)$) to $\pi_{\ell2(\gamma)}$ decay widths provides access to the ratio $| V_{us} / V_{ud} |$~\cite{Marciano:2004uf,FlaviaNetWorkingGrouponKaonDecays:2010lot,Cirigliano:2022yyo}. On the other hand, photon-inclusive $K_{\ell3(\gamma)}$ decays ($K \to \pi \ell \nu_\ell (\gamma)$) are used to determine $|V_{us}|$ directly, ideally using a dispersive representation of the form factors~\cite{Bernard:2007cf,Bernard:2009zm}, together with input on the form-factor normalisation from lattice QCD~\cite{FlavourLatticeAveragingGroupFLAG:2021npn} and isospin-breaking corrections~\cite{Cirigliano:2001mk,Bijnens:2007xa,Cirigliano:2008wn,Seng:2021boy,Seng:2021wcf,Seng:2022wcw}. In contrast, semileptonic $K_{\ell4}$ decays  offer a unique opportunity to probe strong dynamics at low energies: since the final state contains two pions, they are ideal to study $\pi\pi$ interaction~\cite{Shabalin1963,Cabibbo:1965zzb,NA482:2010dug}. In particular, the determination of $\pi\pi$ $S$-wave scattering lengths from $K_{\ell4}$ decays can be compared to very precise theoretical predictions based on Roy equations matched to two-loop ChPT~\cite{Ananthanarayan:2000ht,Colangelo:2000jc} and taking into account important isospin-breaking corrections~\cite{Colangelo:2008sm,Bernard:2013faa}. Furthermore, $K_{\ell4}$ decays are the best source of information about some of the $\mathcal{O}(p^4)$ LECs of $SU(3)$ ChPT.

On the experimental side, impressive precision was reached by the high-statistics measurements of the E865 experiment at BNL~\cite{BNL-E865:2001wfj,Pislak:2003sv} and the NA48/2 experiment at CERN~\cite{NA482:2010dug,NA482:2012cho}. The statistical errors of the $S$-wave of one form factor reached in both experiments the sub-percent level. Matching this precision requires a theoretical treatment beyond one-loop order in the chiral expansion~\cite{Bijnens:1994ie}. Even at two loops~\cite{Amoros:2000mc}, ChPT is not able to predict the observed curvature of one of the form factors.

For the dispersive treatment of the $K_{\ell4}$ form factors of Ref.~\cite{Colangelo:2015kha}, a model-independent parametrisation valid up to and including $\mathcal{O}(p^6)$ was employed, known as reconstruction theorem~\cite{Stern:1993rg,Ananthanarayan:2000cp}. This framework is based on unitarity, analyticity, and crossing, and it includes a resummation of $\pi\pi$- and $K\pi$-rescattering effects. The dispersion relation leads to a coupled system of integral equations, which can be solved numerically using input on the elastic $\pi\pi$- and $K\pi$ scattering phase shifts~\cite{Ananthanarayan:2000ht,Caprini:2011ky,Buettiker:2003pp,Boito:2010me}. The system is parameterised by a few subtraction constants, which are fit to the experimental form-factor data, together with a constraint from the soft-pion theorem~\cite{Callan:1966hu,Weinberg:1966zz}. Isospin-breaking and radiative corrections beyond the ones included in the experimental analyses were computed in Ref.~\cite{Stoffer:2013sfa} at one loop in ChPT including photons and leptons.

The resummed rescattering effects are expected to give the most important contributions beyond $\mathcal{O}(p^6)$ and indeed it turns out that the dispersive description is able to reproduce the experimentally measured form-factor curvature. The dispersion relation enables an analytic continuation of the form factors beyond the physical region and the matching to  ChPT can be performed at zero energy, where the chiral expansion should converge best. The matching to one-loop and two-loop ChPT leads to the results for the LECs $L_1^r$, $L_2^r$, and $L_3^r$ shown in Table~\ref{tab:LECsFromKl4}. The LECs are universal parameters of $SU(3)$ ChPT and enter the description of many mesonic processes at low energies, e.g., (in the case of $L_3^r$) $\eta\to3\pi$~\cite{Colangelo:2018jxw}.

Future experimental improvement on the form factors could reduce further the uncertainties on the LECs $L_1^r$, $L_2^r$, and $L_3^r$. More information on the dependence on the dilepton invariant mass could be used to determine $L_9^r$. Better data could also give access to valuable information on $K\pi$ scattering, and in particular data on the muonic mode $K_{\mu4}$ would provide access to a third form factor that is helicity suppressed and invisible in $K_{e4}$ decays.

\subsection[Kaon experiments as $\pi^0$ factories: a theory point of view]{Kaon experiments as $\boldsymbol{\pi^0}$ factories: a theory point of view }\label{sec:pi0}

Having a primary beam of protons hitting the target, one can get not only kaon flux but naturally also pions. Thus, any typical kaon facility (as today's NA62) would also represent a pion factory. Even if the secondary beam were composed only of kaons (at NA62 the charged kaons represent 6\%), due to the hadronic decay modes of kaons to pions, one would have again a clean source of the lightest mesons. 
For the charged kaons, the dominant hadronic mode is $K^+ \to \pi^+\pi^0$ (its branching ratio is approximately 20.7\%). Due to the very different lifetime of pions ($3\times 10^{-8}$ and $8\times 10^{-17}$s for the charged and neutral pion, respectively), it is unlikely that one can measure both secondary pions by the same detectors. Here, the focus is on the neutral pion decay modes. The $\pi^0$ meson is in some sense unique as it represents the lightest hadron. It plays a crucial role in the study of low-energy properties of the strong interaction and is also important in various scenarios of BSM. There are two fundamental parameters connected with the $\pi^0$ decays -- the pion decay constant $F_\pi$ and the lifetime. $F_\pi$ represents the fundamental order parameter of the spontaneous chiral symmetry breaking. Its standard determination from the pion weak decay relies on the validity of the SM. A tension between its determination from the weak decay $\pi_{l2}$ and the direct $\pi^0$ decay would strongly indicate NP~\cite{Kampf:2009tk,Bernard:2007cf}, although in this comparison the role of isospin-breaking effects, especially the definition of $F_\pi$ in the presence of electromagnetic interactions, becomes critical~\cite{Gasser:2010wz}. The second parameter, the $\pi^0$ lifetime is important for the normalisation of other processes (including kaons).

Let us briefly summarise all observed decay modes with their corresponding branching fractions: $\pi^0\to\gamma\gamma$ ($98.823(34) \times 10^{-2}$), $\pi^0\to\gamma e^+e^-$ ($1.174(35) \times 10^{-2}$), $\pi^0\to e^+e^+e^-e^-$ ($3.34(16) \times 10^{-5}$), $\pi^0\to e^+e^-$ ($6.85(35) \times 10^{-8}$) and $\pi^0\to \gamma \,\text{positronium}$ ($1.82(29) \times 10^{-9}$).
Within the SM (including massive neutrinos), there are further possible decay modes. The pure $\pi^0\to\nu\bar\nu$ is helicity suppressed and, similarly, $\pi^0\to\gamma\nu\bar\nu$ seems to be very far from being measured in present or next-generation experiments. Within the SM, the first process that might be also seen is $\pi^0 \to 4\gamma$ (being roughly 3 orders below the theoretical prediction~\cite{McDonough:1988nf}). It might be a very important process as it goes via the anomalous $\pi^0\gamma\gamma$-vertex and the interesting light-by-light scattering.
Besides, $\pi^0$ decays are also ideal for studying BSM physics~\cite{Moulson:2013oga}, in searches for dark photons~\cite{NA62:2019meo}, for C-parity violation~\cite{McDonough:1988nf,Akdag:2022sbn}, or by a precision measurement of $\pi^0\to e^+e^-$.   

The observed decay modes of neutral pions are governed mainly by the above mentioned $\pi^0\to\gamma^*\gamma^*$ transition form factor $F_{\pi\gamma^*\gamma^*}$, including the dilepton decay $\pi^0\to e^+e^-$ via a loop process. Its knowledge is also important for the anomalous magnetic moment of the muon  $a_\mu$ as it enters hadronic light-by-light (HLbL) scattering via the pion-pole contribution. The uncertainty 
{of}
HLbL scattering is subdominant with respect to hadronic vacuum polarisation, but still relevant, and efforts to reduce the theoretical uncertainty are ongoing~\cite{Aoyama:2020ynm,Colangelo:2022jxc}.   

It is therefore interesting to study and improve our present knowledge of the above $\pi^0$ decay modes. Starting with the dominant $\pi^0\to\gamma\gamma$, one can compare the most precise theoretical estimate based on the EM corrections and the two-loop ChPT     calculation~\cite{Kampf:2009tk}, with the most recent experimental measurement:
\begin{equation}
\text{theory:  }\Gamma
(\pi^0\to\gamma\gamma)=8.09(11) \text{ eV},\qquad
\text{PrimEx:  } \Gamma(\pi^0\to\gamma\gamma)=7.80(12) \text{ eV}
\end{equation}
leading to almost 2$\sigma$ difference. The substantial improvement in recent years from an error of 10\% down to 1.5\% is solely due to the Primakoff-type measurements at JLab \cite{PrimEx-II:2020jwd}. Given the difference with the theory, it is desirable to verify the measurement by a different method, e.g., at a kaon facility.

An even bigger tension was reported for the $\pi^0\to e^+e^-$ decay by the KTeV E799-II experiment \cite{KTeV:2006pwx}. This rare process is important for BSM studies as its long-range SM contribution is loop-induced and chirally suppressed, in such a way that potential BSM effects could compete.  
Further, a discrepancy could have implications for our understanding of $F_{\pi\gamma^*\gamma^*}$ and $g-2$, while at the same time providing valuable BSM constraints.  
However, the original more than $3\sigma$ discrepancy is shifted down to $1.7\sigma$ if radiative corrections are correctly incorporated~\cite{Vasko:2011pi,Husek:2014tna}.  
The problem with the radiative corrections lies in the fact that the extra radiative photon is experimentally  indistinguishable from a Dalitz decay $\pi^0\to\gamma e^+e^-$. Its dedicated study~\cite{Husek:2015sma,Husek:2018qdx} is thus important in the complete understanding of the two-lepton decay mode and the pion transition form factor. With radiative corrections now directly included in the experimental analyses, cf.~Sec.~\ref{sec:MC}, a precision measurement of $\pi^0\to e^+e^-$ can then be confronted with the most recent theoretical study, $\Br(\pi^0\to e^+e^-)=6.25(3)\times 10^{-8}$~\cite{Hoferichter:2021lct}, including $Z$ exchange and relying on detailed dispersive analyses of $F_{\pi\gamma^*\gamma^*}$ in the context of HLbL scattering~\cite{Hoferichter:2014vra,Hoferichter:2018kwz,Hoferichter:2018dmo}, cf.~Sec.~\ref{sec:SM_predictions}, providing a precision test of the SM in analogy to the rare kaon decay $K_L\to \ell^+\ell^-$.  

In summary, the $\pi^0$ decay modes represent a complex and interrelated system important for our understanding of the fundamental interactions that can be studied naturally in future kaon experiments.

\subsection[The time dependent rate $K(t)\to\mu\mu$]{The time dependent rate $\boldsymbol{K(t)\to\mu\mu}$}\label{sec:Grossman}
While improvements for the SM predictions for the $K_{L,S}\to\mu^+\mu^-$ modes are under way (see Secs.~\ref{sec:Kls-mumu}  and~\ref{sec:SM_predictions}), bringing in particular  the  long-distance effects under good control, it is also worthwhile to imagine measuring the time dependent rate, $\Gamma(K\to\mu^+\mu^-)(t)$, from which the theoretically-clean, pure short-distance contribution can be extracted, proportional to the Wolfenstein parameter $\bar\eta$.

The time dependent decay rate of an initial neutral kaon beam is given in terms of the following function of time~\cite{ParticleDataGroup:2022pth} 
\begin{eqnarray}\label{eq:time-dep}
   \frac{1}{{\cal N}}\left(\frac{{\rm d} \Gamma}{{\rm d}t} \right) = 
    f(t)\equiv  \, C_L \,e^{-\Gamma_L t} + C_S \,e^{-\Gamma_S t} +  2\,C_\text{Int.}\cos(\Delta M_K t-\varphi_0) e^{-(\Gamma_L+\Gamma_S)t/2}\, , 
\end{eqnarray}  
where ${\cal N}$ is a normalisation factor, $\Gamma_L$($\Gamma_S$) is the $K_L$($K_S$) decay width, and $\Delta M_K$ is the $K_L\text{--}K_S$ mass difference. 
The four \textit{experimental parameters}, $\{C_L,\,C_S,\, C_{\rm Int.},\, \varphi_0\} ,$
are directly related to the four \textit{theory parameters} describing the system~\cite{Dery:2021mct}, 
\begin{equation}
\left\{ |A(K_S)_{\ell=0}|,\,\, |A(K_L)_{\ell=0}|, \,\, |A(K_S)_{\ell=1}|, \,\,\arg\big[A(K_S)_{\ell=0}^* \, A(K_L)_{\ell=0}\big]\right\}\, ,
\end{equation}
where $\ell=0$ ($S$-wave symmetric wave function) and $\ell=1$ ($P$-wave  anti-symmetric wave function) correspond to the CP-odd and -even $(\mu^+\mu^-)$ final states, respectively. 

Under the following assumptions, all fulfilled to an excellent approximation within the SM --- (i) CP violation  in mixing is negligible, (ii) no scalar leptonic operators are relevant, and (iii) CP violation in the long-distance physics is negligible --- the $\ell=1$ amplitude for $K_L$ vanishes (since it is a CP-odd transition which cannot be induced by vectorial short-distance operators),
$ |A(K_L)_{\ell=1}| \, = \, 0\,$.
Moreover, under the above assumptions the CP-odd amplitude, $|A(K_S)_{\ell=0}|$, is a pure short-distance parameter.

The relations between the experimental and theory parameters can be written as
\begin{align}\label{eq:match-th-exp}
	\begin{aligned}
		C_L\,\,\,\, &= \, |A(K_L)_{\ell=0}|^2\, , \\ 
		C_S\,\,\,\, &= \, |A(K_S)_{\ell=0}|^2 + \beta_\mu^2 |A(K_S)_{\ell=1}|^2 \, , 
		\\ 
		C_\text{Int.}\, &= \, D |A(K_S)_{\ell=0}| |A(K_L)_{\ell=0}| \,, \\
		\varphi_0 \,\,\,\,\, &= \, \arg\big[A(K_S)_{\ell=0}^* \, A(K_L)_{\ell=0}\big]\, ,
	\end{aligned}
\end{align}
where
\begin{equation}
	\beta_\mu = \sqrt{1-\frac{4m_\mu^2}{M_{K}^2}}\,,
\qquad
	D \, = \, \frac{N_{K^0}-N_{\overline{K}^0}}{N_{K^0} +N_{\overline{K}^0}}\, .
\end{equation}
From Eq.~\eqref{eq:match-th-exp}, one can extract the pure short-distance parameter, $|A(K_S)_{\ell=0}|$, using a fit to the experimental parameters (together with knowledge of the dilution factor, $D$),
\begin{eqnarray}\label{eq:extraction}
	\frac{1}{D}\frac{C_{\rm Int.}^2}{ C_L} \, = \, |A(K_S)_{\ell=0}|^2\, .
\end{eqnarray}

In terms of SM CKM parameters, the branching ratio, $\Br(K_S\to\mu^+\mu^-)_{\ell=0}$, is related to the amplitude of interest via
\begin{equation}
	\Br(K_S\to\mu^+\mu^-)_{\ell=0} \, = \, \frac{\tau_S \beta_\mu}{16\pi M_K} |A(K_S)_{\ell=0}|^2\, ,
\end{equation}
and is predicted to an excellent precision.
The prediction, in terms of Wolfenstein parameters, is given by~\cite{Dery:2021mct,Brod:2022khx}
\begin{equation}
	\Br(K_S\to\mu^+\mu^-)_{\ell=0}^{\rm SM}\, = \,  \frac{\tau_S\beta_\mu}{16\pi M_K} \left|\frac{2 G_F^2 M_W^2}{\pi^2}\, f_K M_K m_\mu Y_t \times A^2 \lambda^5 \bar\eta\, \right |^2\, ,
\end{equation}
where $Y_t$ is a loop function, dependent on $x_t = m_t^2/M_W^2$, that is known, and the only hadronic parameter is $f_K$, known to a very high accuracy.
%
The current numeric SM prediction reads~\cite{Brod:2022khx}
\begin{equation}
\label{eq:SMnumeric}
\Br(K_S\to\mu^+\mu^-)_{\ell=0}^{\rm SM}\, = \, 1.70(02)_{\rm QCD/EW}(01)_{f_K}(19)_{\rm param.} \times 10^{-13}\, ,
\end{equation}
where the non-parametric uncertainties are of the order of $\sim 1\%$.

This demonstrates the potential of a future measurement of $\Gamma(K\to\mu^+\mu^-)(t)$, marking it a \textit{third kaon golden mode}. To summarise recent theory progress:
\begin{enumerate}
    \item A measurement sensitive to interference effects in the time dependent $K\to\mu^+\mu^-$ rate can be used to extract the CP-violating short-distance mode, $\Br(K_S\to\mu^+\mu^-)_{\ell=0}$, which provides a clean measurement of 
    $\left|V_{ts}V_{td}\sin(\beta+\beta_s)\right|\approx |A^2 \lambda^5 \bar \eta|$,
	with theory uncertainty of ${\cal O}(1\%)$~\cite{DAmbrosio:2017klp,Dery:2021mct,Brod:2022khx}.
	
    \item The combination of a measurement of $\Br(K_S\to\mu^+\mu^-)_{\ell=0}$ with a measurement of $\Br(K_L\to\pi^0\nu\bar\nu)$ results in a ratio that is a very clean test of the SM, in particular avoiding $|V_{cb}|$-related uncertainties~\cite{Buras:2021nns}.
	
    \item The same $\Br(K_S\to\mu^+\mu^-)_{\ell=0}$ observable is a potent probe of NP scenarios affecting the kaon sector, complementary to the sensitivity of $K_L\to\pi^0\nu\bar\nu$~\cite{Dery:2021vql}.
	
    \item The phase shift characterising the $K_L$--$K_S$ oscillations in the $K(t)\to\mu^+\mu^-$ rate is also cleanly predicted within the SM, up to a fourfold discrete ambiguity~\cite{Dery:2022yqc}.
\end{enumerate}

These recent results are an example for novel experimental and theoretical ideas that are developing, driven by the prospect of the realisation of future kaon factories.

\subsection{Discussion: SM predictions -- continuum}
\label{sec:SM_predictions}
 Theoretical uncertainties in SM predictions for kaon decays can be roughly separated into (i) parametric uncertainties, which can be reduced by improving the precision of CKM input parameters, (ii) perturbative corrections, which can be improved by higher-order loop corrections, see Sec.~\ref{sec:gold-plated theory}, and (iii) long-distance contributions, which are traditionally calculated in ChPT~\cite{Cirigliano:2011ny}. Limitations concern the energy range in which predictions apply and the knowledge of LECs. The latter one is particularly severe for many radiative channels, since, due to gauge invariance, some form factors may receive contributions only at high orders, when LECs parameterising unknown high-energy contributions proliferate, or be dominated by resonance contributions that are only poorly reproduced by the chiral expansion. As discussed in  Sec.~\ref{sec:ChPT}, a promising strategy in this case concerns the interplay with large-$N_c$ considerations and matching to short-distance constraints. 
 
 Moreover, in recent years lattice QCD has made remarkable progress in calculating the respective matrix elements (cf.\ Sec.~\ref{sec:Lattice}), but also improved continuum strategies have been developed that allow one to extend the scope and predictive power of ChPT. One such strategy concerns the use of dispersion relations to unitarise amplitudes,  capturing rescattering effects that are known to impede a rapid convergence of ChPT, e.g., for $S$-wave $\pi\pi$ scattering (cf.\ Sec.~\ref{sec:ChiPT and dispersion}). In addition, dispersive techniques can be used to constrain LECs directly, e.g., in many cases elastic contributions can be calculated unambiguously in terms of known form factors. Recent examples of such ideas include radiative corrections to $K_{\ell 3}$ decays~\cite{Seng:2021boy,
Seng:2021wcf,Seng:2022wcw} (including input from lattice QCD), the kaon mass difference~\cite{Stamen:2022uqh}, and even matrix elements for neutrinoless double-$\beta$ decay~\cite{Cirigliano:2020dmx,Cirigliano:2021qko}. In general, further constraints arise from the short-distance behaviour, where both perturbative calculations and large-$N_c$ arguments may prove useful. In many cases, the different methods are complementary, in such a way that combined analyses can help further improve the precision. 

\begin{figure}[t]
	\centering	\includegraphics[width=0.3\linewidth]{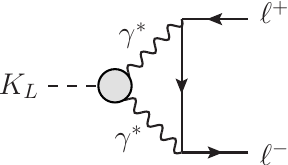}
	\caption{Long-distance contribution to $K_L\to\ell^+\ell^-$. The $K_L\to\gamma^*\gamma^*$ form factor is indicated by the gray blob.} 
	\label{fig:KLmumu}
\end{figure} 

As a concrete example, 
some of these techniques have been used to improve the long-distance contribution to $K_L\to\ell^+\ell^-$, which arises from the $K_L\to\gamma^*\gamma^*$ form factor as shown in Fig.~\ref{fig:KLmumu} and needs to be controlled if constraints on the short-distance amplitude~\cite{Buchalla:1993wq,Gorbahn:2006bm} are to be extracted. In one-loop ChPT, the point-like form factor in the diagram generates a UV divergence, which becomes absorbed by a LEC with unknown finite part~\cite{GomezDumm:1998gw}. Moreover, the full dynamics of the $K_L\to\gamma^*\gamma^*$ form factor are only resolved at subleading orders, at which yet new parameters appear. However,  similar form factors have been studied with dispersive techniques in great detail in the context of the HLbL contribution to the anomalous magnetic moment of the muon~\cite{Aoyama:2020ynm}, in the case of $\pi^0$~\cite{Hoferichter:2018dmo,Hoferichter:2018kwz,Hoferichter:2021lct}, $\eta^{(')}$~\cite{Holz:2015tcg} and $f_1$~\cite{Zanke:2021wiq,Hoferichter:2023tgp}, and similar strategies apply to the $K_L$~\cite{Hoferichter:2023wiy}. Including data from both leptonic ($K_L\to\ell^+\ell^-\gamma$) and hadronic ($K_L\to\pi^+\pi^-\gamma$) processes and matching to the asymptotic contribution~\cite{Isidori:2003ts} in terms of a dispersive integral indeed allows one to reduce the uncertainty of the resulting prediction for the long-distance part of the $K_L\to\ell^+\ell^-$ amplitude. Further improvements, including a definite statement on the relative sign between long-distance and short-distance contributions, should become possible in combination with input from lattice QCD~\cite{Zhao:2022pbs,ChaoPos2023,Hoid:2023has}, see Sec.~\ref{sec:Kls-mumu}. Finally, continuum, data-driven techniques profit from data on related processes, e.g., in the case of $K_L\to\ell^+\ell^-$ a significant part of the error budget in Ref.~\cite{Hoferichter:2023wiy} derives from experimental uncertainties in the spectra of $K_L\to\ell^+\ell^-\gamma$ and $K_L\to\pi^+\pi^-\gamma$, which could thus be reduced with improved measurements at future kaon facilities. 


\section{Kaon physics beyond the Standard Model}
\label{sec:BSM}
The SM provides a successful and economical description of particle physics up to energies of about 1~TeV. However, there are various phenomenological and theoretical reasons that motivate an extension of this theory at higher energies.  The EW hierarchy problem (i.e., the instability of the Higgs potential under quantum corrections) and the unexplained hierarchies of the SM Yukawa coupling (the so-called flavour problem) are among the most compelling theoretical arguments in favour of new (heavy) degrees of freedom. In this perspective, the SM can be viewed as the renormalisable part of an effective field theory valid up to some still undetermined cutoff scale $\Lambda$. There are no direct indications about the value of $\Lambda$; however, natural solutions of the hierarchy problem suggest that it should not exceed a few TeV.   
\subsection{The BSM potential of rare kaon decays}
\label{sec:BSM_potential}

Indirect NP searches, such as those conducted via FCNC processes, aim at probing the SM cutoff by looking at suppressed SM processes, where the relative impact of new degrees of freedom can be larger. In this perspective, rare $K$ decays, and in particular the theoretically clean  $K\to\pi\nu\bar\nu$ modes,  play a unique role, since they could allow us to  probe, for the first time and with high precision, the short-distance structure of the $s\to d$ FCNC amplitude.

The $s\to d$ transition is very interesting since it experiences a twofold suppression within the SM: 1) it is forbidden at the tree-level, and 2) it is further strongly suppressed by the hierarchy of the SM Yukawa couplings. In all kaon decays measured so far it is impossible to get precise short-distance information about the $s\to d$ transition (i.e., determining the strength of this transition at the EW scale): the interesting short-distance dynamics is obscured by large long-distance effects. This does not happen only in $K\to\pi\nu\bar\nu$ decays. This is the reason why these processes are i) very interesting, ii) very suppressed, and iii) can be predicted with high accuracy in the SM. 

\begin{figure}[t]	
  \hskip -0.5 cm
 \includegraphics[width=0.5\linewidth]{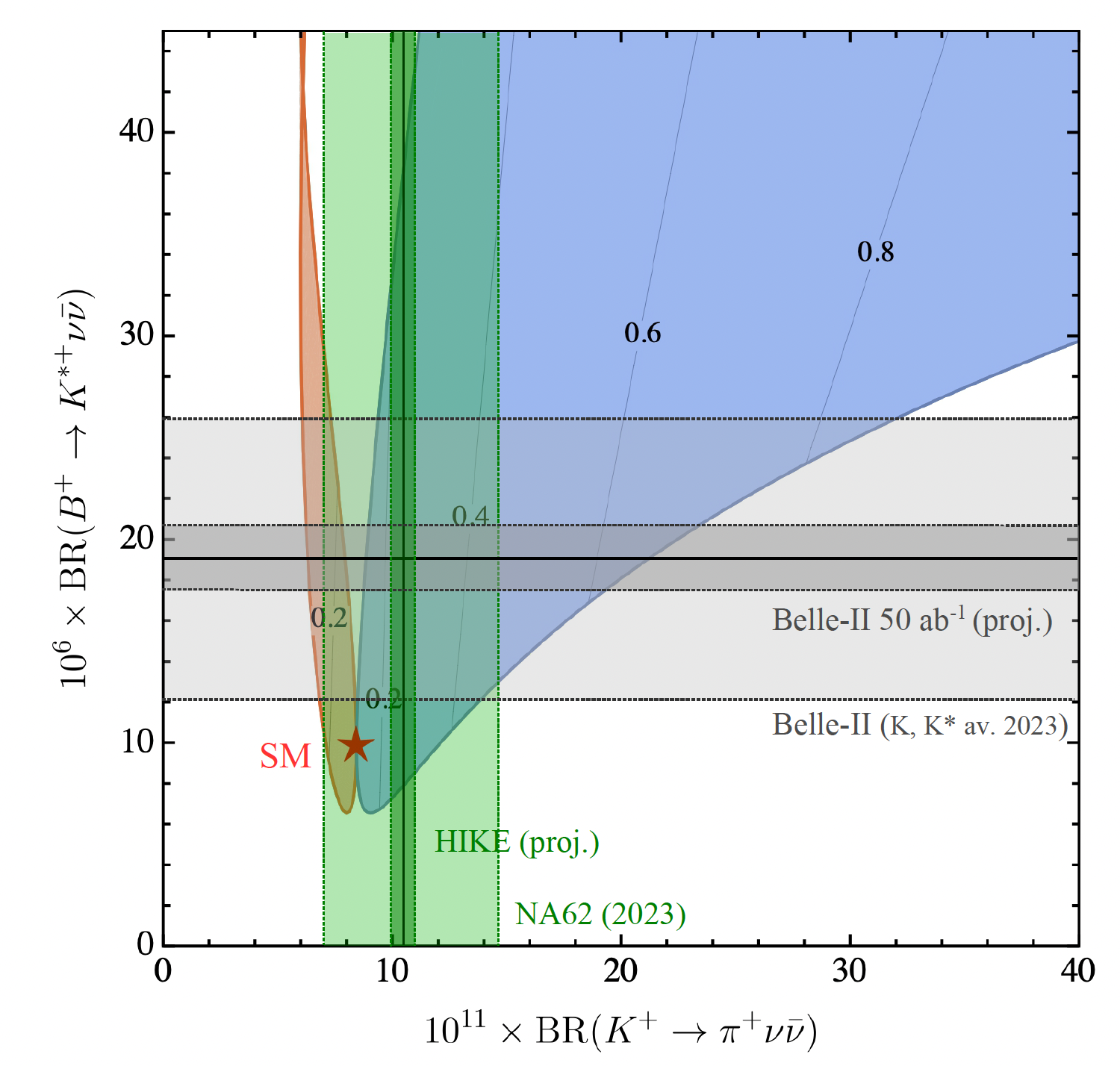} 
 \hskip -0.2 cm
 \includegraphics[width=0.52\linewidth]{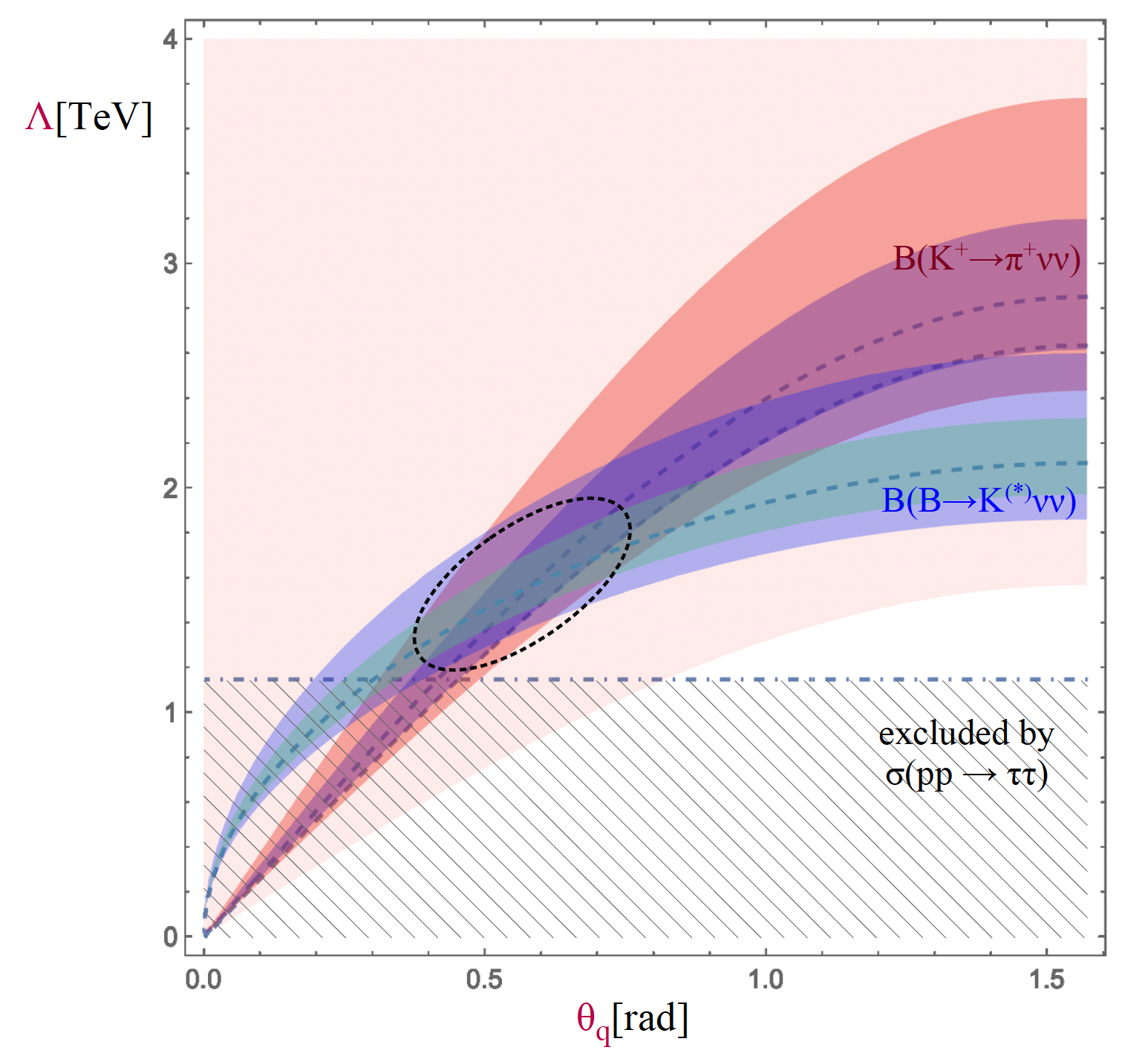}
	\caption{
 $\Br(K^+ \to \pi^+ \nu\bar \nu)$ vs.~$\Br(B \to K^{(*)} \nu\bar \nu)$ in extensions of the SM with NP coupled 
 dominantly to the $3^{\rm rd}$ family leading to an effective interaction of the type  $(\bar q_L^3 \gamma^\mu T q_L^3) (\bar\ell_L^3 \gamma_\mu T \ell_L^3)$, with $T=\sigma^a$ (triplet) or $T=1$ (singlet).
 Left: allowed values for the two modes, 
 setting $R_{D}= 1.25$, and assuming a 
 triplet interaction (red)
 or triplet - 2~singlet (blue); 
 the bands denote present errors and future projections~\cite{Bordone:2017lsy}.
 Right: scale of the effective operator (singlet case) vs.~the parameter $\theta_q$ describing the flavour alignment 
 [$q^3_L\equiv \cos(\theta_q) b_L +\sin(\theta_q) t_L$]; 
 the coloured bands denote $1\sigma$ and 
 2$\sigma$ regions assuming a 
 $5\%$ [$10\%$] measurement of 
 $\Br(K^+ \to \pi^+ \nu\bar \nu)$ 
 [$\Br(B \to K^{(*)} \nu\bar \nu)$] around the present central value; 
 the pink area indicates the present allowed region.
 These plots highlight the importance of combining precise data on the two modes in 
 determining not only the effective scale of NP, 
 but also its EW and  flavour structure.
 } 
	\label{fig:BSM}
\end{figure}

$\Br(K^+ \to \pi^+ \nu\bar \nu)$  and $\Br(K_L \to \pi^0 \nu\bar  \nu)$ are conceptually comparable to clean precise EW observables, such as the $W$ mass or the Higgs self-coupling. Similarly to those, rare $K$ decays 
probe EW dynamics. However, rare $K$ decays probe it in a different and less tested sector connected to the flavour problem, which is one of the main reasons why the SM needs to be extended. Not surprisingly, there are plenty of examples in the literature of well motivated BSM models, perfectly consistent with present high-energy data, which would give rise to large deviations from the SM in both decay modes (see, e.g., Refs.~\cite{Straub:2013zca,Buras:2015yca,Ishiwata:2015cga,Bordone:2017lsy,Fajfer:2018bfj,Mandal:2019gff,Marzocca:2021miv,Crosas:2022quq}).
A further unique aspect of  $K\to\pi\nu\bar\nu$  decays is that they are sensitive to the interaction of light quarks ($s$ and $d$) with third-generation leptons (the $\tau$ neutrinos). This additional unique aspect enhances their sensitivity to motivated BSM models shedding light on the origin of the flavour hierarchies (see, e.g., Refs.~\cite{Bordone:2017lsy,Crosas:2022quq,Davighi:2023iks}).
Note that, besides being particularly motivated, such models are also favoured by present data that indicates 
an excess, over the SM predictions, 
in $b\to c\tau\nu$ decays. An illustration of the potential impact of a future precise measurement of $\Br(K^+ \to \pi^+ \nu\bar \nu)$ in this context, also in connection with the expected precision on $\Br(B \to K^{(*)} \nu\bar \nu)$,
is shown in Fig.~\ref{fig:BSM}.

Loosely speaking, the motivated BSM theories that can be tested by means of  $K\to \pi\nu \bar\nu$ decays fall into the same category as those researched at HL-LHC, i.e., theories with new heavy particles in the few TeV regime. An explicit example of the complementary of $K\to \pi\nu \bar\nu$ 
with collider and electroweak observables 
in this context
has recently been shown in~\cite{Allwicher:2023shc}.
In this respect, it is worth emphasising that NP around the TeV scale is perfectly compatible with current data,
and still represents the most natural option to address the EW hierarchy problem~\cite{Allwicher:2023shc}. 
Limits on specific exotic particles, or new contact interactions, can even exceed 100~TeV, but this fact should not be over-interpreted: these strong limits only apply to exotic states that badly violate some of the accidental symmetries of the SM.  Low-scale NP models compatible with current data imply small deviations from the SM and can only be searched for via dedicated precision measurements. 

As with all precision tests, also in the case of $\Br(K\to \pi\nu \bar\nu)$ the impact of the result depends  both on the experimental and on the theoretical accuracy. Right now, the experimental precision on $\Br(K^+ \to \pi^+ \nu\bar \nu)$ is around 40\%, well above the theoretical one. 
Reaching the few-percent level on this mode would represent major progress. If, as predicted in motivated NP models, a significant deviation from the SM were observed, this would be a major breakthrough. However, the value of this observable is such that a precision measurement would have far-reaching implications even in case of a result compatible with the SM. This would allow us to rule out (or further constrain) a class of well motivated BSM theories in the TeV-energy domain addressing the origin of the flavour hierarchies.

\subsection{Exotica from kaon decays: theory and experiment }\label{sec:exotica}
The question of how to rank searches for exotics or forbidden decays inevitably runs into theoretical prejudices. One can for instance order NP models in terms of their simplicity such as the minimal field content that is added to the existing SM structure. One can see that this may not be the best criterion by considering that before the discovery of the muon or even charm, simplicity would not have guided one to the complex structure of the SM. The other option is to ask for interesting experimental signatures, so that no stone is left unturned. This can easily lead to rather complicated model building (``signature building''), and also runs against the problem of limited resources. 

An option that is quite appealing is to give priority to the models that already solve some of the outstanding problems in particle physics. An example of such a model is the QCD axion, which solves both the strong CP problem and is a cold dark matter candidate. In this case HIKE is in a very advantageous position that it will probe very interesting parameter space. Assuming sizeable flavour-violating couplings, such as in the case of axiflavon \cite{Calibbi:2016hwq},  HIKE will probe values of axion decay constant, $f_a$, that result in the correct DM abundance in minimal scenarios. 

\subsubsection{Theory overview}
Heavy NP with flavour-violating couplings is very efficiently probed via classic FCNC probes such as $K$--$\bar K$ mixing, $\mu \to e$ conversion, rare $B$ decays, etc. The effects of such heavy NP are encoded in the values of the Wilson coefficients, for instance $C_{ds} (\bar d \gamma^\mu s)(\bar d \gamma_\mu s)$ for $K$--$\bar K$ mixing. The tree level exchanges of heavy NP mediators will result in $C_{ds}\propto g^2/M^2$, where $g$ is the flavour-violating coupling of the heavy mediator to the quark current, $M$ the mediator mass, and we are omitting ${\mathcal O}(1)$ factors. That is, the classic FCNC probes with SM particles as asymptotic states can probe very heavy mediators that have  ${\mathcal O}(1)$ flavour-violating couplings, or much lighter NP that is proportionally more weakly coupled. 

There is a qualitative change in the sensitivity, if one searches for rare decays to light NP states. For instance, if light NP state $\phi$ couples through dimension 5 operators that are suppressed by $1/f_a$, the branching ratio for $K\to \pi \phi$ is parametrically given by $\Br(K\to \pi \phi)\propto (M_W^2/f_a M_K)^2$. That is, the sensitivity to the UV scale $f_a$ is parametrically enhanced because the kaons are light. The underlying reason is that the kaon decay width is suppressed by the off-shellness of the $W$, so that $\Gamma_K\propto M_K^5/M_W^4$. This same heuristic argument applies to other light mesons and leptons: which one wins depends on flavour and CP structure of NP.  

The other open question is how ``exotic'' are such light NP states. Here, an important comment is that any global $U(1)$ that is spontaneously broken will result in a light (pseudo-)\linebreak Nambu--Goldstone boson. The perhaps most celebrated example is the QCD axion, while in general such light pseudoscalars $a$ go under the term axion-like particles or ALPs. These can have flavour-violating couplings \cite{Wilczek:1982rv,Davidson:1981zd,Feng:1997tn, Kamenik:2011vy,Bjorkeroth:2018dzu,MartinCamalich:2020dfe}. If these are already present in the UV then $K\to \pi a$ decays probe very high scales, $f_a\sim 10^{13}$ GeV, while if the flavour violation is due to the SM CKM, generated at 1-loop, then the reach is correspondingly lower, at $f_a\sim 10^6$ GeV \cite{Bauer:2021mvw,Goudzovski:2022vbt}. Beyond ALPs, there are many other well motivated models of light NP to which kaon decays are sensitive, such as light Higgs-mixed scalar, heavy neutral leptons, dark photon, etc. Even seemingly exotic signatures could in fact be due to relatively simple extensions of the SM. One such example, is for instance the decay $K\to \pi 2(e^+e^-)$ that would be generated within a dark Higgsed $U(1)_d$, where both the Higgs and the dark photon are light. A more complete list can be found in Ref.~\cite{Goudzovski:2022vbt}.


\subsubsection{Principal experimental signatures}

Discussed below are the principal $K^+$ decay channels that have been exploited by NA62, and will be exploited at HIKE, to address the Physics Beyond Collider (PBC) benchmark scenarios (BC) in the classification of Ref.~\cite{Beacham:2019nyx}.  The updated HIKE sensitivity is published in the HIKE proposal~\cite{HIKE-Proposal}.

Kaon decays also provide sensitivity to a large number of non-minimal scenarios that evade detection in beam-dump experiments~\cite{Harris:2022vnx,Gori:2022vri,Ballett:2019pyw}, which have been studied experimentally  only to a minimal extent so far. Examples of non-minimal scenarios accessible in kaon experiments are: short-lived Majorana heavy neutral leptons (HNLs) decaying via a displaced-vertex topology $K^+\to\ell_1^+N$, $N\to\pi^-\ell_2^+$~\cite{Atre:2009rg}; dark neutrino produced and decaying via the $K^+\to\ell^+N$, $N\to\nu Z'$, $Z'\to e^+e^-$ chain~\cite{Ballett:2019pyw}; a muonphilic force scenario leading to $K^+\to\mu^+\nu X$ decays~\cite{Krnjaic:2019rsv}.


\boldmath
\paragraph{$\boldsymbol{K^+\to\pi^+X_{\rm inv}}$ decays:}
\unboldmath
The search for the $K^+\to\pi^+X_{\rm inv}$ decay, where $X_{\rm inv}$ is an invisible particle, provides sensitivity to the benchmark scenarios BC4 (dark scalar), BC10 (ALP with fermion coupling) and BC11 (ALP with gluon coupling). The accessible $m_X$ ranges are approximately 0--110~MeV/$c^2$ and 150--260~MeV/$c^2$, corresponding to the $K^+\to\pi^+\nu\bar\nu$ signal regions. The principal background comes from the $K^+\to\pi^+\nu\bar\nu$ decay itself. The search strategy based on the peak search in the spectrum of the reconstructed missing mass $m_{\rm miss}^2=(P_{K^{+}}-P_{\pi^{+}})^{2}$ has been established by the NA62 experiment~\cite{NA62:2020xlg}, and the full NA62 Run~1 (2016--2018) dataset has been analysed~\cite{NA62:2021zjw}. The search at HIKE Phase~1 will be performed by direct extension of the $K^+\to\pi^+\nu\bar\nu$ measurement. The HIKE Phase~1 sensitivity projection has been performed by extension of the NA62 analysis, assuming a 40-fold increase in the size of the data sample with respect to NA62 Run~1.

The above scenarios are also addressed by a dedicated search for the $\pi^0\to X_{\rm inv}$ decay using a technique established by NA62~\cite{NA62:2020pwi}. This search covers the $m_X$ region in the vicinity of the $\pi^0$ mass. The region $m_X>260~{\rm MeV}/c$ for the scenarios BC4 and BC11 is addressed experimentally by searches for $K^+\to\pi^+X$ decays followed by displaced $X\to\mu^+\mu^-$ or $X\to\gamma\gamma$ decays, respectively.


\boldmath
\paragraph{$\boldsymbol{K^+\to\ell^+N}$ decays:}
\unboldmath
Searches for the $K^+\to\ell^+N$ decays ($\ell=e,\mu$), where $N$ is an invisible particle, provide sensitivity to the benchmark scenarios BC6 (HNL with electron coupling) and BC7 (HNL with muon coupling). The technique has been established by the NA62 experiment, which has obtained world-leading exclusion limits on the HNL mixing parameters $|U_{\ell 4}|^2$ over much of the accessible mass range of 144--462~MeV/$c^2$ with the Run~1 dataset~\cite{NA62:2020mcv,NA62:2021bji}. Both searches are limited by background. In particular, the $K^+\to\mu^+\nu$ decay followed by $\mu^+\to e^+\nu\bar\nu$ decay in flight, and the $\pi^+\to e^+\nu$ decay of the pions in the unseparated beam, represent irreducible backgrounds to the $K^+\to e^+N$ process.

The HIKE sensitivity projection is obtained by extension of the NA62 analysis assuming the similar resolution and background. In the $K^+\to\mu^+N$ case, it is assumed additionally that, unlike NA62, the trigger line is not downscaled, which is possible for a fully software trigger. HIKE sensitivity to $|U_{e4}|^2$ in the mass range 144--462~MeV/$c^2$ approaches the seesaw neutrino mass models~\cite{Abdullahi:2022jlv}. For $m_N<140~{\rm MeV}/c^2$, HIKE will improve the PIENU limits~\cite{PIENU:2017wbj} on $|U_{e4}|^2$ via the $\pi^+\to e^+N$ decays of pions in the unseparated beam, and has a further potential via the $K^+\to\pi^0e^+N$ decay~\cite{Tastet:2020tzh}. HIKE will also approach the seesaw neutrino mass models for $|U_{\mu4}|^2$.


\boldmath
\paragraph{$\boldsymbol{\pi^0\to\gamma A'}$ decay:}
\unboldmath
A search for the $K^+\to\pi^+\pi^0$, $\pi^0\to\gamma A^\prime$, $A'\to e^+e^-$ prompt decay chain had been performed by the NA48/2 experiment~\cite{NA482:2015wmo}, addressing the benchmark dark photon scenario BC1. The case of invisible dark photon (scenario BC2) has been addressed by the NA62 experiment~\cite{NA62:2019meo}. The HIKE experiment will be able to improve on both searches. Of particular interest for the future experimental programme are the displaced $A^\prime\to e^+e^-$ vertex analysis which potentially provides sensitivity for lower dark photon couplings, and a study of an alternative dark photon production channel $K^+\to\mu^+\nu A'$, followed by either prompt or displaced $A'\to e^+e^-$ decays, extending the search region above the $\pi^0$ mass.


\paragraph{Other processes:}
Other exotic processes studied recently using the NA62 Run~1 dataset include searches for lepton-flavour and -number violating decays, including $K^+\to\pi^-\mu^+\mu^+$ and $K^+\to\pi^-e^+e^+$~\cite{NA62:2019eax}, $K^+\to\pi^\pm\mu^\mp e^+$ and $\pi^0\to\mu^-e^+$~\cite{NA62:2021zxl}, $K^+\to\pi^-(\pi^0)e^+e^+$~\cite{NA62:2022tte}, and $K^+\to\mu^-\nu e^+e^+$~\cite{NA62:2022exp}. These searches are almost background-free, and typically reach sensitivities to the decay branching ratios of ${\cal O}(10^{-11})$. The sensitivities will be improved significantly with NA62 Run~2 and HIKE datasets.

The NA62 experiment has recently reported the first search for pair-production of hidden-sector mediators in the prompt $K^+\to\pi^+aa$, $a\to e^+e^-$ and $K^+\to\pi^+S$, $S\to A'A'$, $A'\to e^+e^-$ decay chains leading to a five-track final state~\cite{NA62:2023rvm}.

\subsection{Discussion: (B)SM Constraints from kaon physics }\label{sec:bsm-dicuss}

In this section the impact of kaon observables as
constraints on the SM and beyond is discussed.

The interplay of kaon physics with other areas of particle physics can
be nicely illustrated by the example of anomalous couplings of the $Z$
boson to top quarks. These couplings can be measured directly at the
LHC via $t\bar t + Z$ production~\cite{Rontsch:2014cca}. It turns out,
however, that indirect constraints are more
powerful~\cite{Brod:2014hsa}.  In particular, rare $B$ and $K$ meson
decays are sensitive probes of such couplings. Moreover, anomalous
$t\bar t Z$ couplings are related, via gauge invariance, to anomalous
couplings of $W$ bosons, which leads to a rich interplay between rare
decays, EW precision observables, and collider signals.

\begin{figure}[h]
 \centering
 \includegraphics[width=\textwidth]{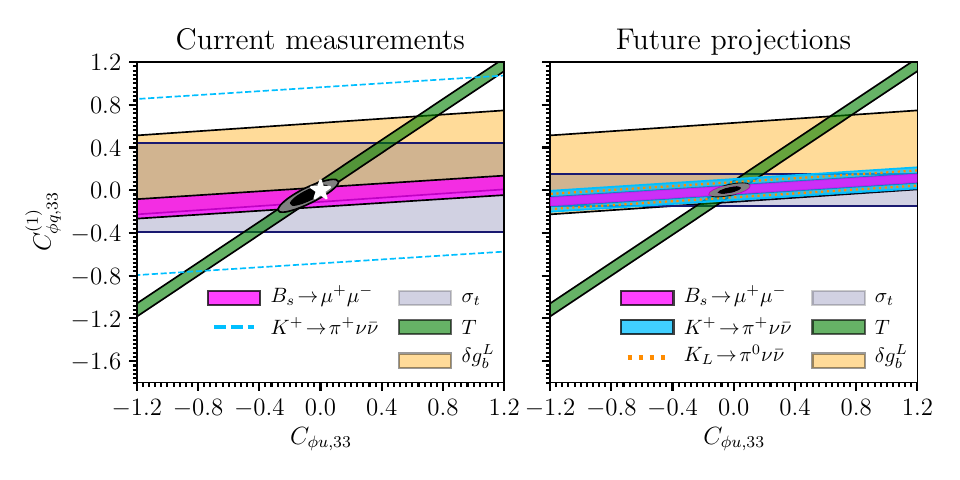}
 \caption{Constraints on anomalous $t\bar tZ$ couplings. The left panel shows current constraints on the two independent coefficients $C_{\phi q,33}^{(1)}$ and $C_{\phi u,33}$, arising from the rare decays $B_s \to \mu^+ \mu^-$ and $K^+ \to \pi^+ \nu \bar \nu$, $t$-channel single top production ($\sigma_t$), as well as the EW precision parameters $T$ and $\delta g_b^L$. The white star indicates the SM expectation. The right panel shows future projections, assuming $5\%$ measurements of the rare-decay modes and a naive rescaling of the uncertainty in single top production with $300/\text{fb}$ of data. The EW precision parameters were measured at LEP and are not expected to change significantly. See text for discussion.
   \label{fig:ttZ}}
\end{figure}

This interplay can be discussed in a transparent way in the context of
SM effective field theory (SMEFT)~\cite{Buchmuller:1985jz, Grzadkowski:2010es}. At mass dimension
six, there are three operators that induce anomalous $t\bar t + Z$
couplings at tree level, namely, $Q_{\phi q,33}^{(3)}$,
$Q_{\phi q,33}^{(1)}$, $Q_{\phi u,33}$. Strong constraints from $B$
meson decays require the combination
$C_{\phi q,33}^{(3)} + C_{\phi q,33}^{(1)}$ to be very small. In
certain models with vector-like quarks, this combination vanishes
identically~\cite{delAguila:2000rc}, which is assumed in the
following. In this case, rare $K$ and $B$ meson decays put strong
bounds on non-standard $t\bar t Z$ couplings. Currently, the
$B_s \to \mu^+ \mu^-$ mode is clearly dominant (Fig.~\ref{fig:ttZ},
left panel), but precise measurements of the $K \to \pi \nu \bar \nu$
will lead to comparable (and complementary) constraints in the
future. The single other most important constraint arises from the
measurement of the $T$ parameter, while the modification of the
left-handed $b\bar b Z$ (i.e.,  $\delta g_b^L$) leads to weaker
constraints. The modification of the charged current affects the
$t$-channel single top production cross section which has been
measured by both ATLAS~\cite{ATLAS:2023hul} and
CMS~\cite{CMS:2020vac}; however, rare decays are expected to give
stronger bounds even in the future with a larger LHC data sample.

Rare kaon decays are also sensitive probes of light new particles that
can appear in the final state. For instance, the measurement of the
$K^+ \to \pi^+ \nu \bar \nu$ branching ratio can set stringent bounds
on well-motivated models of axion dark matter if the axions have
flavour off-diagonal couplings, see the discussion in
Sec.~\ref{sec:exotica} for details.

The expected $K^+ \to \pi^+ \nu \bar \nu$ event rate at HIKE will result in a measurement of the invariant mass spectrum of the final-state neutrino pair and test fundamental properties of neutrinos.
If we consider only lepton-number-conserving interactions of SMEFT, the resulting three light Majorana neutrinos can only couple through an axial vector current with the quark sector in the limit of small neutrino masses.
The resulting missing mass spectrum would be a rescaled SM spectrum.
Dimension seven operators, that violate lepton number, can generate scalar interactions and (neutrino flavour-changing) tensor interactions.
Neutrino flavour-conserving scalar interactions were studied in Ref.~\cite{Deppisch:2020oyx} and a sensitivity to energy scales in the multi TeV range was found.
The neutrino spectrum peaks at higher invariant masses if compared with the SM expectation.
The situation changes if we go beyond the three light Majorana neutrino scenario.
Three extra neutrinos can potentially form Dirac neutrinos, together with the lepton doublets of the SM.
In this scenario, scalar and tensor interactions are already generated at dimension six in $\nu$SMEFT without the need for lepton number violation.
{Preliminary results \cite{Gorbahn:2023juq, GMSSTinprep} for this scenario and extra sterile Majorana neutrinos were discussed.}
Both scenarios are sensitive to NP in the 100\,TeV range.
Dirac neutrinos also allow for lepton-flavour-conserving tensor interactions.
The scenario with extra massive sterile neutrinos can lead to unique modifications of the spectrum through  the modified phase space.

\subsection[Discussion: Complementarity of $B$- and $K$-decays]{Discussion: Complementarity of $\boldsymbol{B}$- and $\boldsymbol{K}$-decays}
Flavour physics is a traditional source of correlation among $B$-and $K$-physics: are there observables and models correlating $B$- and $K$-physics? What about lepton-flavour-universality violation  (LFUV)?   Which models? What about model independent tests?  CP violation? Which LFUV scale can be tested? What about lepton-flavour-violating (LFV) decays like $K\to \pi \mu e$?

Based on the stronger constraints of FCNCs of the first two families compared to the third family, the traditional Minimal Flavour Violation (MFV) protection of FCNCs based on $U(3)$ flavour symmetry was already challenged by a less protective $U(2)$: indeed Isidori and collaborators have recently applied this $U(2)$ to interesting $B$- and $K$- correlations  \cite{Barbieri:2015yvd,Marzocca:2021miv}, in particular $B\to \pi \nu \bar{\nu}$ and $K\to \pi \nu \bar{\nu}$. Also a typical NP scale of $\sim \ 1\text{--}2\; {\rm TeV} $ was indicated.

Since a SMEFT approach generically predicts LFV in higher dimensional operators, it is interesting to question LFUV and LFV in kaon physics independently, and also address which kaon observable could be more interesting; for instance, LFUV was tested in  $K^\pm\to \pi ^\pm  \ell^+ \ell^- $ \cite{Crivellin:2016vjc}: in the presence of LFUV in higher dimensional operators, for instance $Q_{7V}$, affecting differently $K^\pm\to \pi ^\pm  e^+ e^- $ and $K^\pm\to \pi ^\pm  \mu^+ \mu ^- $,  experiment could test LFUV by measuring  the form factors of the  different final states.   

However, if NP affects left-handed currents also neutrinos of different flavours are affected in $K\to \pi \nu \bar{\nu}$. Moreover, $K_L \to \mu^+ \mu ^- $ and $K_L\to\pi^0\ell^+\ell^-$ experiments may give relevant constraints to LFUV (dimension-six $V-A\otimes V-A$ operators)
coefficients \cite{DAmbrosio:2022kvb}.  In this research all kaon decays potentially constraining LFUV interactions
($K_{L}\to \pi ^{0} \nu \bar\nu$, $K^+\to \pi^+ \nu \bar\nu$, $K_L\to\mu^+\mu ^- $ and $K_L\to\pi^0\ell^+\ell^-$) are limiting the supposed  different  couplings to the first and second left-handed family. Also the projections for the future programmes of KOTO and NA62, i.e. KOTO-II and HIKE Phases~1 and 2, are studied.

One of the questions of this workshop  was how to quantify the impact of BSM reach, through dimension-6 operators, in flavour and collider experiments (kaons, beauty, $\Delta F=2,1 $, muon decays, dipole moments, Higgs decays, \ldots): one can parameterise  the limits in terms of the scale appearing  
in the  Wilson coefficients of dimension-6 operators, but it was also argued  that specific models could be more useful to show the experimental reach. For instance there are specific  models that might be more effective to address $B$- and  $K$-physics and the possible $g-2$ muon anomaly. Other studies have shown  that the explanation of $B\to D^{(*)}\tau \bar{\nu}$  decay would generate, through $W$-box diagrams, effects in $K\to \pi \nu \bar{\nu} $~\cite{Crivellin:2018yvo}.

Several theoretical and experimental studies have explored the possibility of lepton number violation manifesting itself in rare kaon decays. Also NA62, KOTO, KOTO-II and HIKE have studied this, see Sec.~\ref{sec:exp}.
Several models have been  discussed at the workshop: i) using SMEFT   its detection would put high-scale leptogenesis under tension and would hint to small radiatively generated neutrino masses~\cite{Deppisch:2020oyx},
ii) German Valencia and collaborators~\cite{He:2022ljo} compare constraints on pairs of light scalars or vectors from their contribution to  $K\to \pi \nu \bar{\nu} $ and $B \to  M \nu \bar{\nu} $ $(M = K, \pi$ etc). 

One interesting question in the workshop was the possible connection between the strong CP problem and the flavour problem: it was discussed the possibility  that solving the flavour problem of the SM with a simple $U(1)_H$ flavour symmetry naturally leads to an axion that solves the strong CP problem and constitutes a viable Dark Matter candidate. This framework is very predictive and experimentally testable by future axion and precision flavour experiments \cite{Calibbi:2016hwq}.

The interplay between kaon physics and other areas is also evident in precision measurements of charged-current decays of $K^+$ and $K_L$. These decays serve as a robust BSM probe through the testing of first-row CKM unitarity and the exploration for nonstandard currents. They are sensitive to effective TeV scales, complementing EW precision data and direct LHC searches, as can be seen in the SMEFT framework~\cite{Cirigliano:2009wk,FlaviaNetWorkingGrouponKaonDecays:2010lot,Gonzalez-Alonso:2016etj,Cirigliano:2023nol}. The current discrepancies observed across various modes provide further motivation to delve into these processes and resolve the existing uncertainties.


\section{Outlook and Conclusions}
\label{sec:conclusions}
This document provides a compact summary of talks and discussions from the workshop Kaons@CERN 23~\cite{KaonsatCERN}. 
Over 100 leaders in experiment and theory participated in the workshop to take stock, discuss and contemplate about the opportunities that current and future kaon-physics
experiments, as well as anticipated theoretical developments, provide for particle physics in the coming decade and beyond. A few outcomes are worth highlighting:
\begin{itemize}
\item The rare-decay channels $K^+\to\pi^+\nu\bar\nu$ and $K_L\to\pi^0\nu\bar\nu$ are among the theoretically cleanest standard candles of the SM. Being essentially free of hadronic uncertainties they 
 allow for accurate and precise tests of the SM. Their suppression in the SM leads to sensitivity to NP at the highest scales.
\item The theoretical precision matching the projections of future experiments HIKE and KOTO-II does already exist. 
\item Generic NP models show complementarity of searches in $B$ and $K$ decays. Results from a future kaon factory will uniquely impact the constraining or understanding of the microscopic structure of NP.
\item \begin{sloppypar}Besides the gold-plated rare modes, HIKE and KOTO-II will measure a plethora of other $K^+$ and $K_L$ decays (rare, less rare and radiative) with unprecedented precision. For some of these decays the improved measurements will help the community of SM theorists working in ChPT, dispersion theory and lattice QCD+QED to test their predictions and sharpen their tools.\end{sloppypar}
\item
For several other decays not belonging to the gold-plated class, the improved measurements will help put further constraints on NP models, in particular if one analyses them in combined fits. Moreover their effectiveness can increase if theoretical calculations of long-distance contributions improve, even at a later stage. Precision HIKE measurements of the dominant $K^+$ and $K_L$ decay modes can also provide important NP input.
\item While the best possible outcome is that NP will be discovered by HIKE and/or KOTO-II, the resulting precision measurement of the SM are guaranteed deliverables and will allow for stronger constraints of NP models, and hence, stronger exclusion limits.
\end{itemize}
As the workshop has shown -- the kaon physics community is diverse, young and vibrant and distributed around the entire world. 
A clear commitment to a kaon factory in terms of HIKE will give it a further boost and allow it to further develop ideas and projects that are already ongoing. 
Historically, the study of kaon physics was an important driver for the development of the SM, but it is far from a closed chapter: kaon physics still harbours many fundamental questions. As this workshop has clearly highlighted, through their sensitivity to high-scale physics, kaons could very well 
be the place where first signs of NP will be discovered. Studying kaons to high precision with a next-generation kaon factory should therefore be a priority for both CERN and J-PARC.

\vskip 0.3cm

{\bf Acknowledgments:} We acknowledge the kind financial support by LHCb, NA62 and the CERN Department of Theoretical Physics.
\bibliographystyle{jhep}
\bibliography{biblio}
\end{document}

%% file: experimental.tex
%
\section{Experimental kaon physics}
\label{sec:exp}
The three major experiments performing kaon physics are: the NA62 fixed-target decay-in-flight experiment at the CERN north area working with a $K^+$ beam, the KOTO experiment at the J-PARC Hadron Experimental Facility (HEF) working with a $K_L$ beam, and the LHCb experiment at the Large Hadron Collider at CERN with particular sensitivity to $K_S$ and hyperons.

At the CERN north area the HIKE programme is proposed to continue fixed-target decay-in-flight experiments, with much increased beam intensity and a new detector setup, with both $K^{+}$ and then $K_{L}$ beams in a multi-phase project.

At J-PARC, KOTO-II, the evolution of KOTO, is being discussed as a part of the HEF extension project, to reach the sensitivity required to detect tens of $K_{L}\rightarrow\pi^{0}\nu\bar{\nu}$ decays.

The LHCb experiment has already undergone a major upgrade, which includes a paradigm shift to using a software trigger. 
This is crucial for the $K_S$ and hyperon programme since previously hardware triggers, not designed for kaon studies, were highly inefficient, while now software triggers can be developed to fully exploit the high luminosity available.

\subsection{Dedicated kaon experiments at CERN}

\subsubsection{The NA62 experiment: present status}
The main aim of the NA62 experiment is the precise measurement of the ultra-rare decay $K^+\to\pi^+ \nu\bar\nu$ using a decay-in-flight technique.
NA62 exploits the CERN SPS 400~GeV/$c$ primary proton beam, that impinges on a beryllium target and produces a 75~GeV/$c$ secondary beam made of positively charged particles of which approximately 6\% are $K^+$. 
The experimental signature of a $K^+ \rightarrow \pi^+ \nu\bar\nu$ decay is an incoming $K^+$ and an outgoing $\pi^+$ with missing energy in the final state. 
The signal is kinematically discriminated from other kaon decays using the squared missing mass $m_{\rm miss} = (p_K - p_{\pi})^2$ variable, where $p_K$ and $p_{\pi}$ are the 4-momenta of the kaon and of the downstream charged particle respectively, in the pion mass hypothesis.

The experiment has taken data in 2016--2018 (Run~1)~\cite{NA62:2017rwk}. NA62 recorded about $3\times 10^{18}$ protons on target in Run~1, and at least twice this value is expected in Run~2.
The analysis of Run~1 data led to the observation of 20 $K^+\to\pi^+\nu\bar\nu$ signal candidates (with about 10 SM signal and $7.03^{+1.05}_{-0.85}$ background events expected).
The measured branching ratio $\Br(K^+ \to\pi^+\nu\bar\nu) = (10.6^{+4.0}_{-3.4}|_\text{stat} \pm0.9_\text{syst}) \times 10^{-11}$ is compatible with the SM prediction within one standard deviation, and corresponds to a significance of $3.4 \sigma$~\cite{NA62:2021zjw}. This is the most precise measurement of the $K^+ \rightarrow \pi^+ \nu\bar\nu$ branching ratio to date and provides the strongest evidence so far for the existence of this extremely rare process.

Data taking has resumed with Run~2 in 2021, and is approved until long shutdown 3 (LS3). Several detector upgrades have been implemented during LS2. A first preliminary analysis of the data collected in 2022 exhibits a sensitivity similar to that of the whole Run~1. A measurement of $\Br(K^+\to\pi^+ \nu\bar\nu)$ with a precision of 15\% is achievable by LS3, assuming a beam delivery similar to that of 2022 in the upcoming years~\cite{NA62-report}.

\begin{figure}[h]
 \centering
 \includegraphics[width=0.97\textwidth]{figs/HIKE_detector_phase1.pdf}
 \includegraphics[width=0.97\textwidth]{figs/HIKE_detector_phase2.pdf}
 \caption{HIKE Phase 1 (top) and Phase 2 (bottom) layouts, with an aspect ratio of 1:10.
   \label{fig:hike_detector}}
\end{figure}

\subsubsection{HIKE Phase 1 and 2: Experimental design and physics reach} 

HIKE is a world-leading comprehensive programme of kaon-decay experiments, which follows a staged approach and includes several phases~\cite{HIKE-Proposal}. The programme focuses on  several ultra-rare ``golden'' kaon-decay modes, which are very clean from the theory point of view (providing unique closure tests of the CKM paradigm), and are exceptionally difficult to measure. The primary goal of Phase~1 is a measurement of the  $K^+\to\pi^+\nu\bar\nu$ branching ratio to a 5\% precision (matching the precision of the SM calculation), while the main goal of Phase~2 is the first observation of the $K_L\to\pi^0\ell^+\ell^-$ decays with a significance above $5\sigma$. The experimental layout for both phases is shown in Fig.~\ref{fig:hike_detector}. HIKE will use $K^+$ and $K_L$ beams of record intensity, and will therefore collect the world's largest samples of $K^+$ and $K_L$ decays using a flexible software trigger and more performant detectors with respect to NA62 and other previous experiments. As a result, HIKE will significantly improve the precision of measurements over a wide range of kaon decay channels, providing unique results of long-lasting scientific value. Table~\ref{tab:hike-flavour} lists a selection of the many unique measurements that HIKE can perform.  
The challenges to be addressed in the HIKE detectors are in synergy with or go beyond to current efforts for LHC-experiment upgrades after Long Shutdown 4, and will help in making a significant step towards the needs for FCC detectors. Further in the future, a third phase would address $K_L \rightarrow \pi^0 \nu\nu$. 

\begin{figure}[h]
 \centering
 \includegraphics[width=0.8\textwidth]{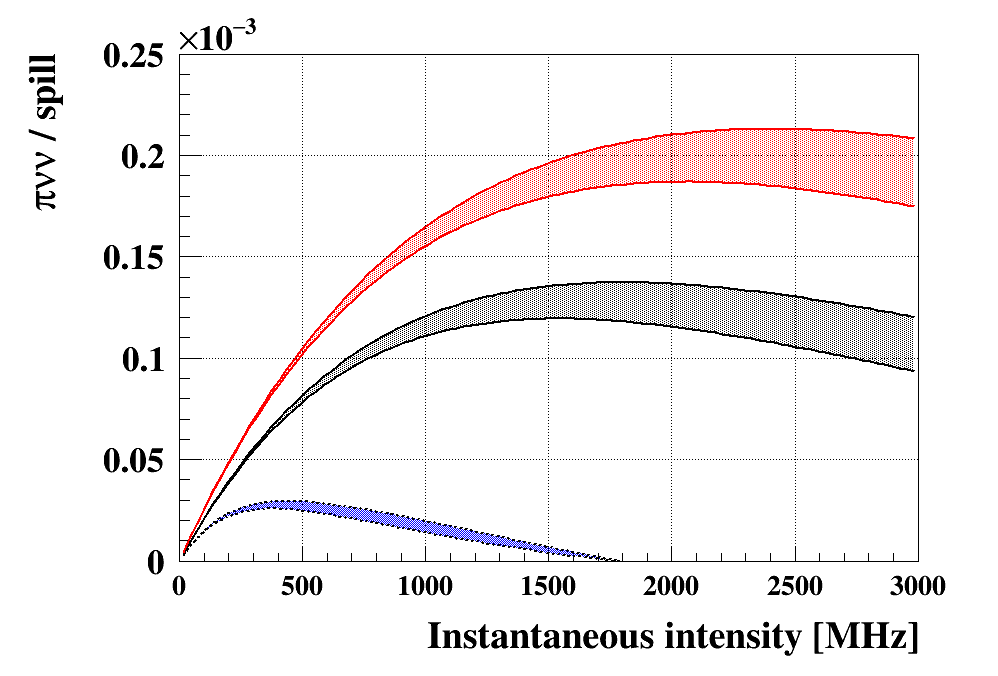}
 \caption{Numbers of selected $K^+\to\pi^+\nu\bar{\nu}$ events per spill as a function of the instantaneous
beam intensity. The blue shaded area shows the number of events from a data-driven model of
the NA62 signal yield. The black shaded area represents the same model but with detector time
resolutions improved by a factor of 4 with respect to NA62, assuming also a software trigger. The
red shaded area represents the final HIKE Phase 1 signal yield model with all improvements included. The
width of the shaded areas illustrates the uncertainty in the intensity dependence model.
   \label{fig:hike_yield}}
\end{figure}

\begin{table}[htb]
{\small
\centering
\caption{Summary of HIKE sensitivity for flavour observables. The $K^+$ decay measurements will be made in Phase~1, and the $K_L$ decay measurements in Phase~2. The symbol $\cal B$ denotes the decay branching ratios.}
\begin{tabular}{lll}
\hline
$K^+\to\pi^+\nu\bar\nu$ & $\sigma_{\cal B}/{\cal B}\sim5\%$ & BSM physics, LFUV \\
$K^+\to\pi^+\ell^+\ell^-$ & Sub-\% precision & LFUV \\
& on form-factors\\
$K^+\to\pi^-\ell^+\ell^+$, $K^+\to\pi\mu e$ & Sensitivity ${\cal O}(10^{-13})$ & LFV / LNV \\
Semileptonic $K^+$ decays & $\sigma_{\cal B}/{\cal B}\sim0.1\%$ & $V_{us}$, CKM unitarity \\
$R_K={\cal B}(K^+\to e^+\nu)/{\cal B}(K^+\to\mu^+\nu)$ & $\sigma(R_K)/R_K\sim {\cal O}(0.1\%)$ & LFUV \\
Ancillary $K^+$ decays & \% -- \textperthousand & Chiral parameters (LECs) \\
(e.g. $K^+\to\pi^+\gamma\gamma$, $K^+\to\pi^+\pi^0 e^+e^-$) \\
\hline
$K_L\to\pi^0\ell^+\ell^-$ & $\sigma_{\cal B}/{\cal B} < 20\%$ & ${\rm Im}\lambda_t$ to 20\% precision, \\
&& BSM physics, LFUV \\
$K_L\to\mu^+\mu^-$ & $\sigma_{\cal B}/{\cal B}\sim 1\%$ & Ancillary for $K\to\mu\mu$ physics \\
$K_L\to\pi^0(\pi^0)\mu^\pm e^\mp$ & Sensitivity ${\cal O}(10^{-12})$ & LFV \\
Semileptonic $K_L$ decays & $\sigma_{\cal B}/{\cal B}\sim 0.1\%$ & $V_{us}$, CKM unitarity \\
Ancillary $K_L$ decays & \% -- \textperthousand & Chiral parameters (LECs), \\
(e.g. $K_L\to\gamma\gamma$, $K_L\to\pi^0\gamma\gamma$) & & SM $K_L\to\mu\mu$, $K_L\to\pi^0\ell^+\ell^-$\\
\hline
\end{tabular}
\label{tab:hike-flavour}
}
\end{table}

The primary objective for the first phase of HIKE will be measuring the branching ratio of $K^{+}\rightarrow\pi^{+}\nu\bar{\nu}$ with about $5\%$ precision, improving by a factor of approximately 3 on the NA62 projected precision of about $15\%$ when using the full NA62 dataset.
The statistics required to reach the HIKE Phase 1 goal, which corresponds to about 9 times the NA62 one, will be collected in about 4 years thanks to an increase of a factor 4 in the beam intensity and an increase in the signal acceptance of a factor $>2$ thanks to new, more granular/performant detectors.
To be able to stand the intensity increase, the timing for all the detectors needs to be improved by at least a factor 4 (see Fig.~\ref{fig:hike_yield}). Despite the higher rate, other key performances such as kinematic rejection, photon rejection, and particle identification efficiency must at least be kept equal to the NA62 ones to maintain background rejection under control.
In addition to the precise measurement of its branching ratio, the increased statistics will allow investigating the nature of the $K^{+}\rightarrow\pi^{+}\nu\bar{\nu}$ decay, i.e., vector (SM) vs.\ scalar or tensor (BSM) contributions, that implies testing the fundamental nature of neutrinos.

HIKE precision measurements of other $K^+$ rare decays
will allow studies of the kinematic distributions and form factors with unprecedented precision. 
For $K^{+}\rightarrow\pi^{+}\ell^{+}\ell^{-}$ decays at least $5\times10^{5}$ events in both the $\ell=\mu$ and $\ell=e$ channels will be collected allowing a precision lepton-flavour-universality test and decay spectra
(such as those in the equivalent $B$-physics channel, with a complementary physics reach). 
For $K^{+}\rightarrow\pi^{+}\gamma\gamma$ a branching-ratio precision of a few per-mille will be achieved, to match a similar theory expected precision. 
In addition, precision studies of chiral pertubation theory (ChPT) predictions can be performed 
investigating the details of the $\gamma\gamma$ spectrum, including the near-the-cusp effect, to extract low-energy constants (LECs) (cf. Secs.~\ref{sec:ChPT} and~\ref{sec:ChiPT and dispersion}). 

HIKE Phase~2 will use a $K_{L}$ beam and will allow a general-purpose investigation of $K_{L}$ decays, especially those with charged particles in the final state - exploiting precision tracking and 
particle identification systems, which will mostly be maintained from HIKE Phase~1. The primary goal of HIKE Phase~2 will be $K_{L}\to\pi^{0}\ell^{+}\ell^{-}$, however many other modes will be measured, and in general can only be studied at HIKE Phase~2. The study of $K_{L}\to\pi^{0}\ell^{+}\ell^{-}$ will allow its observation for the first time, and then its measurement
with at least 20\% precision. This decay gives unique access to short-distance BSM effects in the photon coupling via the tau loop~\cite{isidori-seminar}, as well as giving access to the CKM CP-violating parameter $\eta$.

Across HIKE Phases~1 and 2, precision measurements of the most common $K^{+}$ and $K_{L}$ decays can allow new global fits to be performed which will be able to clarify the current Cabibbo anomaly tensions. 
Searches for heavy neutral leptons can also be performed at HIKE, in both phases, reaching 1--2 orders of magnitude better sensitivity than NA62, including searches for dark neutrinos, which will reach the see-saw line.
Besides, searches for the $K^+\to\ell^+N$ decay can measure the coupling directly: while the analysis of beam-dump data depends on the assumption of decay couplings and needs benchmarks for its interpretation, the 
production studies are benchmark independent.

A full investigation of a range of feebly interacting particles (FIPs) will be performed at HIKE, during both standard kaon data-taking and dump-mode operation, where all the benchmarks set by the Physics Beyond Colliders initiative will be investigated~\cite{HIKE-Proposal}, with the exception of BC3.
The most promising channels for FIPs searches in kaon mode are: 
$K^{+}\to\pi^{+}X$, $K_{L}\to\pi^{0}X$, $K\to\pi\pi X$,
where $X$ can be a dark scalar (e.g., the BC4 model), an axion-like particle (ALP) (BC9, BC10) or, if it is very light, an axiflavon;
$K^{+}\to\ell^+N$ where $\ell = e,\mu$ and $N$ is a heavy neutral lepton;
and $K^{+}\to\pi^{+}\pi^{0}$ followed by $\pi^{0}\to\gamma A^{\prime}$ where $A^{\prime}$ is a dark photon (BC1, BC2).
In addition studies of $K^{+}\to\pi^{+}\gamma\gamma$, with a guaranteed SM physics measurement outcome, can also be used to search for BSM physics, in this case scanning 
the $m_{\gamma\gamma}$ invariant mass spectrum to search for evidence of an ALP decaying to two photons, $X\to\gamma\gamma$ (see also the theory 
contributions in Sec.~\ref{sec:exotica}).

Further details of the physics case and experimental setup can be found here~\cite{HIKE-Proposal}.

\subsection{Dedicated kaon experiments at J-PARC}
\subsubsection{The KOTO experiment: present status}
The KOTO experiment at the J-PARC 30 GeV Main Ring is dedicated to the search for the rare decay $K^{0}_{L}\rightarrow \pi^{0} \nu \bar{\nu}$.
This mode directly breaks  CP symmetry and is highly suppressed in the SM. 
In addition, the theoretical uncertainty of this decay is only a few percent. 
These features make this decay one of the best probes to search for NP beyond the SM.
However, due to experimental difficulties, only an upper limit of $3.0\times10^{-9}$ is set by the KOTO experiment with the 2015 data set~\cite{KOTO:2019prk}.

In the analysis of data taken in 2016--18, three signal candidate events were observed with an expected background of $1.22\pm0.26$ events.
The number of observed events was statistically consistent with the background expectation~\cite{KOTO:2020prk}.
The main contribution was the charged kaon contamination in the neutral beam and halo $K^{0}_{L}\rightarrow 2 \pi^{0}$ events, where
$K_{L}$ mesons are scattered at the surface of the collimators and enter the decay region with a large angle. 
Those background events were newly revealed in the analysis. 
A new charged veto counter called UCV was installed in 2021 to detect charged kaons and develop new analysis methods to reduce
the halo  $K^{0}_{L}\rightarrow 2 \pi^{0}$ background. 

The latest analysis is focused on the 2021 data set, 
because the UCV reduces $K^{\pm}$ background (BG) events by a factor of~13 with a 97\% signal efficiency. The halo background is also reduced by a factor of~8 with 92\% signal efficiency by analysis methods newly implemented. Therefore, the numbers of those BG events
are reduced to be less than~0.1. 
Several other analysis methods were implemented to estimate background events more accurately.

\begin{figure}[htbp]
\begin{center}
\includegraphics[width=9cm,angle=-90]{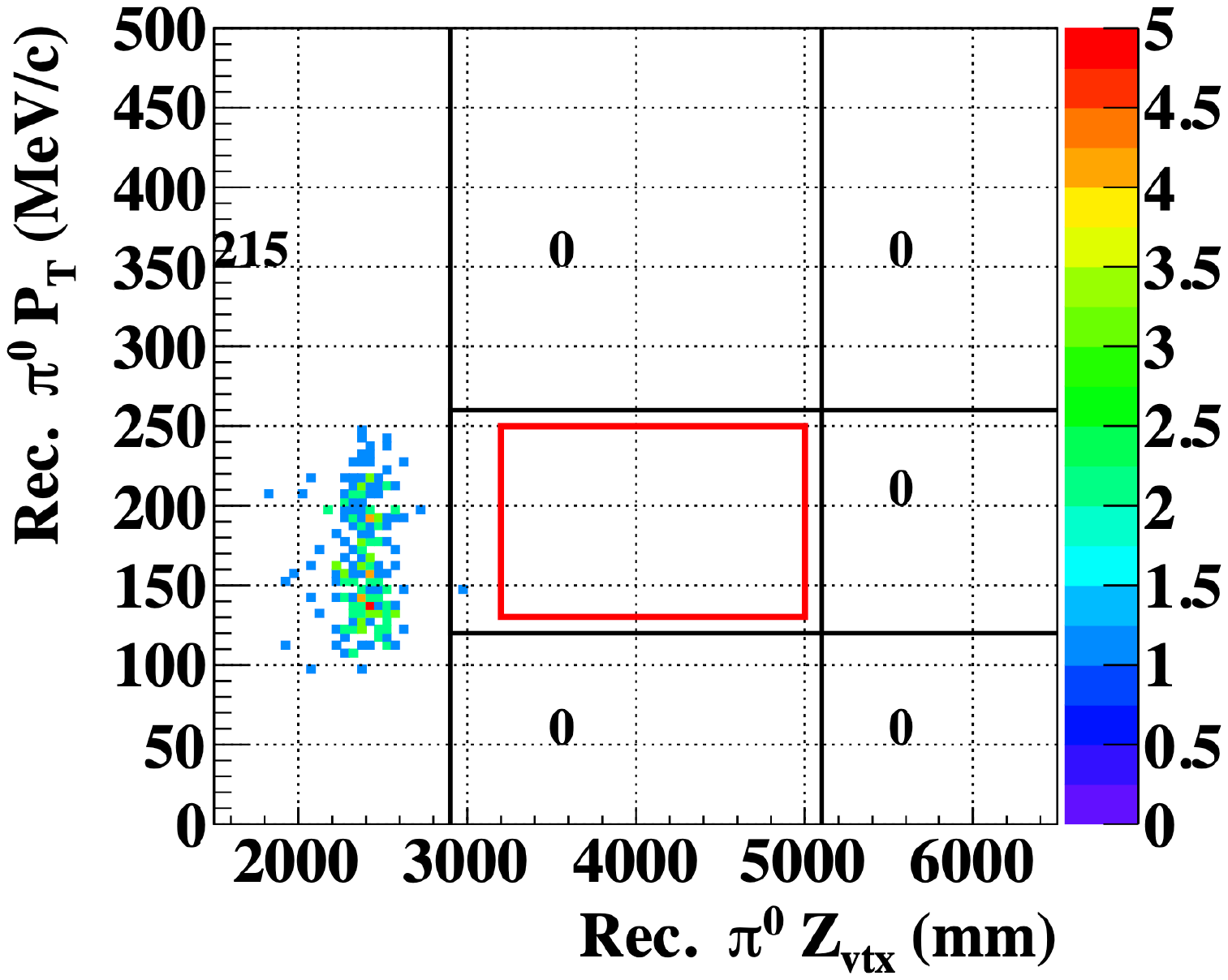}
\caption{Scatter plot of reconstructed P$_{T}$ vs Z$_{vtx}$ for the
2021 data set. The region inside the red line is the signal region. }
\label{fig:PtZ}
\end{center}
\end{figure}

The single event sensitivity (SES) of the 2021 data analysis is $8.7\times10^{-10}$ while the SES of the previous analysis was $7.2\times10^{-9}$.
Table~\ref{tab:BGSummary} summarises the numbers  of the background events expected in the signal box. The largest contribution comes from the upstream $\pi^{0}$ background events, 
where a $\pi^{0}$ is generated by neutrons in the beam halo region in a detector located at the upstream region. 
The second largest contribution comes from the $K^{0}_{L}\rightarrow 2 \pi^{0}$ background events. The total number of BG events expected in the signal box is estimated 
to be $0.255\pm0.058^{+0.053}_{-0.068}$. 
Figure~\ref{fig:PtZ} shows the scatter plot of reconstructed P$_{T}$ vs Z$_{vtx}$ for the
2021 data set: the region inside the red line is the signal region.
No candidate event was observed in the signal region.
An upper limit is therefore set on the branching ratio of the $K^{0}_{L}\rightarrow \pi^{0} \nu \bar{\nu}$ decay to be $2.0\times10^{-9}$ at 90\% C.L. with Poisson statistics. This latest result was presented at the workshop.

KOTO still has 2019--2020 data already collected but not yet finalised in the analysis. Measures to reduce the $K^{\pm}$ background events are needed for this sample,
because only a prototype detector of UCV was present in 2020 and there was no detector to detect $K^{\pm}$ in 2019.
For the future run, KOTO plans to collect 10 times more protons on target (POT) in 4--5 years to achieve a sensitivity below $10^{-10}$.

\begin{table}
	\caption{Preliminary summary of the numbers of background events in the signal region for the 2021 data.}
	\label{tab:BGSummary}
	\centering
	\begin{tabular}{lc}
		\toprule
		source &  Number of events\\
		\midrule
		 Upstream-$\pi^0$ 				  & 0.064 $\pm$ 0.050 $\pm$ 0.006 \\
		 $K_{L}\rightarrow2\pi^{0}$ 		  & 0.060 $\pm$ 0.022 $^{+0.051}_{-0.060}$ \\ 
		 $K^{\pm}$					  & 0.043$\pm$ 0.015 $^{+0.004}_{-0.030}$ \\
		Hadron-cluster		 			  & 0.024$\pm$ 0.004  $\pm$ 0.006 \\
		Halo $K_{L}\rightarrow2\gamma$	  & 0.022$\pm$ 0.005  $\pm$ 0.004 \\
		Scattered $K_{L}\rightarrow2\gamma$  & 0.018$\pm$ 0.007  $\pm$ 0.004 \\
		$\eta$ production in CV 			 & 0.023$\pm$ 0.010  $\pm$ 0.006 \\
		
		\midrule
		total 							& 0.255$\pm$ 0.058 $^{+0.053}_{-0.068}$ \\
		\bottomrule
	\end{tabular}
\end{table}

\subsubsection{KOTO~II prospects and plans} 
\begin{figure}[h]
\centering
\includegraphics[bb=0 0 585 364,clip,width=0.5\textwidth]{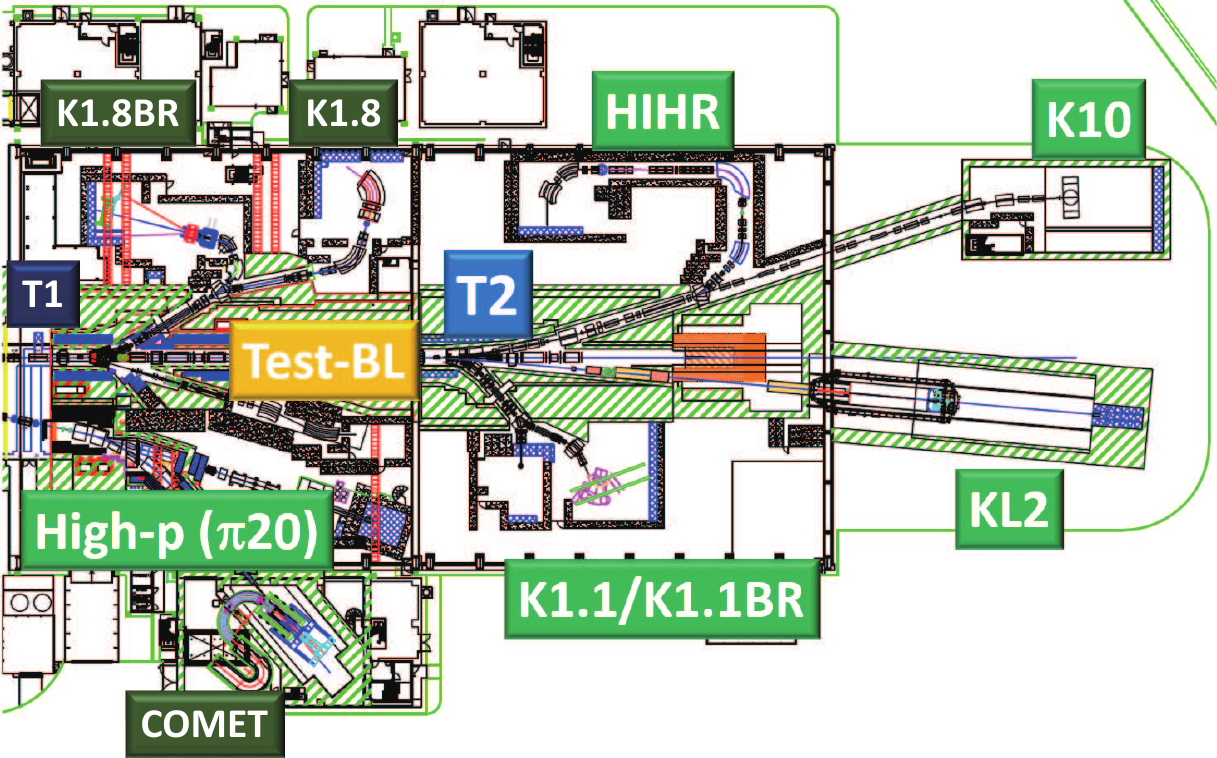}
\caption{Layout of the extended Hadron Experimental Facility at J-PARC. The KOTO~II detector is behind the dump downstream of the T2 target.
\label{fig:HEFex}}
\end{figure}

The KOTO-II experiment, planned at the extended Hadron Experimental Facility of J-PARC, is designed to measure the branching ratio of the decay $K_L\to\pi^0\nu\bar\nu$ (Fig.~\ref{fig:HEFex}). 

The $K_L$ mesons produced at the T2 target are guided to the KOTO-II detector behind the dump
with a 43-m long beamline including two collimators and two magnets.
The extraction angle of $K_L$ is $5^\circ$ with a solid angle of $4.8~\mu\mathrm{sr}$.
The long beamline is designed to reduce short-lived particles;
the two magnets sweep charged particles out, and the
two collimators are designed to suppress beam-halo particles.

\begin{figure}[tb]
\centering
\begin{minipage}{0.65\textwidth}
\includegraphics[bb=14 21 1853 376,clip,width=\textwidth]{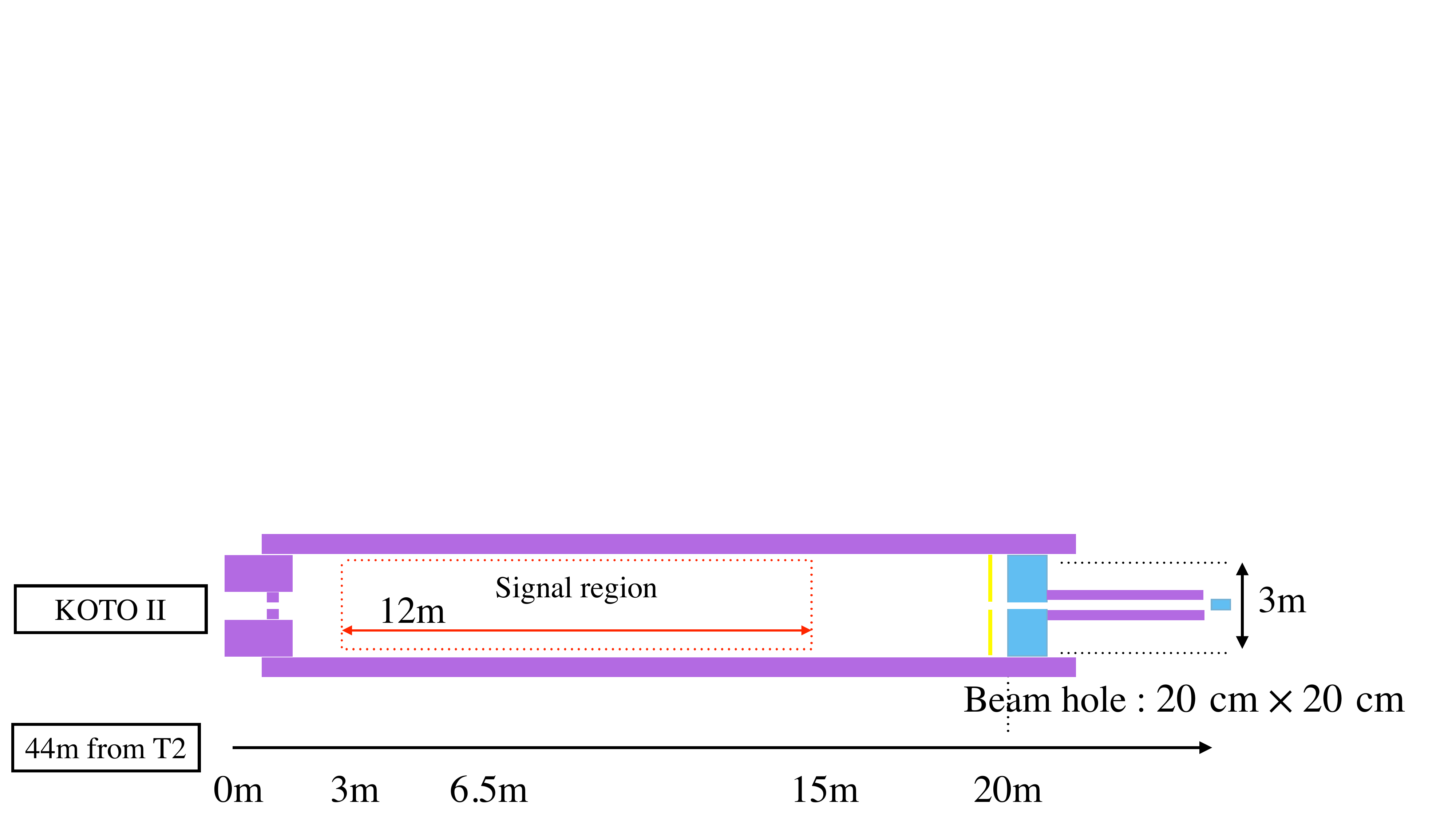}
\caption{Conceptual design of the KOTO~II detector.\label{fig:KOTOIIdetector} }
\end{minipage}
\hspace{0.05\textwidth}
\begin{minipage}{0.25\textwidth}
\includegraphics[bb=0 0 1822 1080,clip,width=\textwidth]{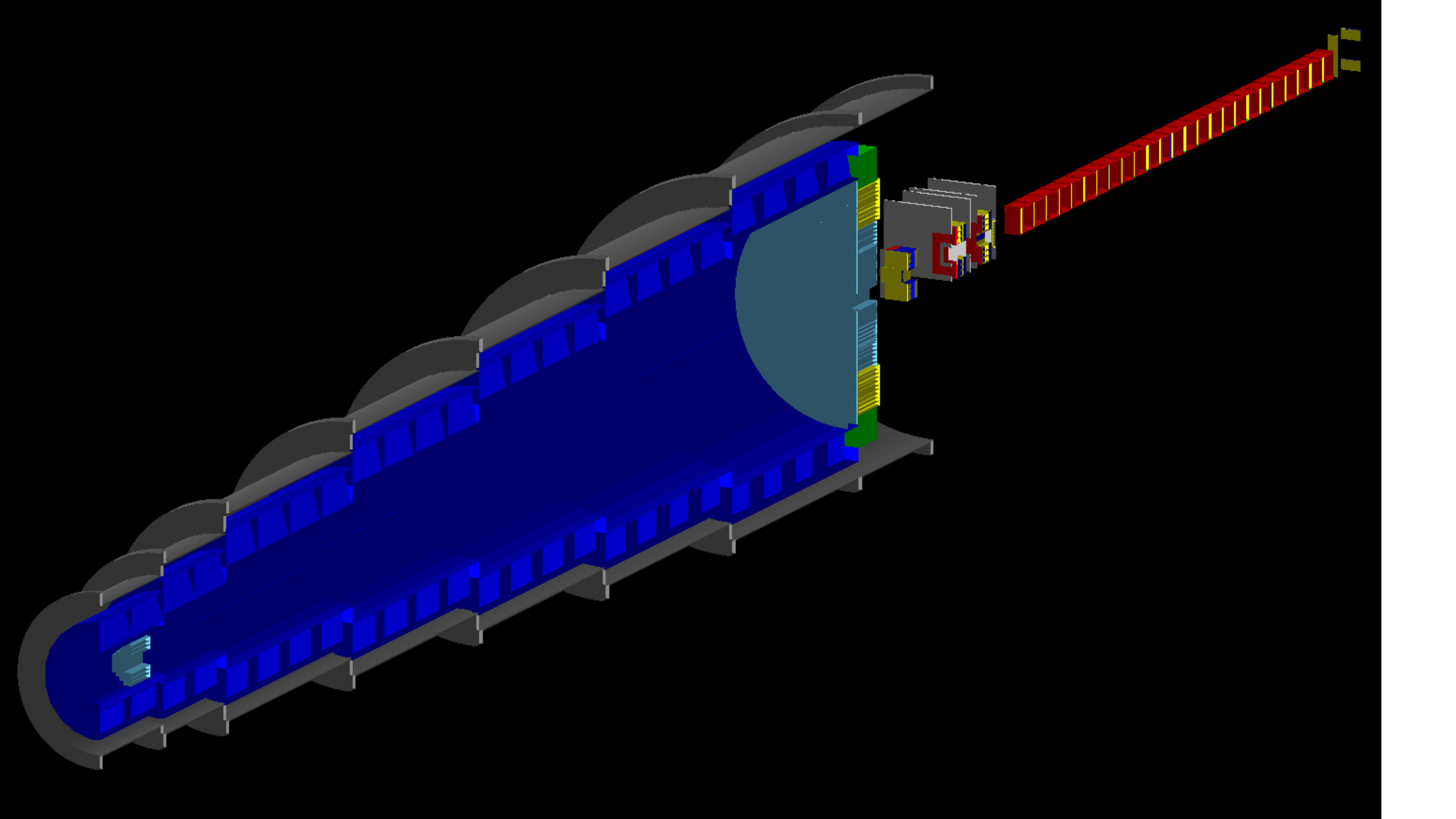}
\caption{Cutaway view of realistic KOTO-II detector.\label{fig:det3d}}
\end{minipage}
\end{figure}

The KOTO~II detector  (Fig.~\ref{fig:KOTOIIdetector})
starts at 44 m from the T2 target,
which is the origin of the axis system.
The $z$-axis is along the beam axis pointing downstream.
The signal decay region is defined by $3 < z < 15 $ meters.
An electromagnetic calorimeter is 3~m in diameter and located at $z=20~\mathrm{m}$.
Veto counters surround the decay region hermetically.
Two photons from the $\pi^0$ in the signal $K_L\to\pi^0\nu\bar\nu$ process are detected with the calorimeter.
The decay vertex of the $\pi^0$ is reconstructed on the $z$-axis
assuming the invariant mass of the two photons to be the nominal $\pi^0$ mass.
The transverse momentum of the $\pi^0$ ($p_T$) is reconstructed using the vertex position.
Events are vetoed if extra particles are detected
other than the two photons from the $\pi^0$ in the calorimeter. 

\begin{figure}[tb]
\centering
\includegraphics[bb=9 33 1474 1045,clip,width=0.9\textwidth]{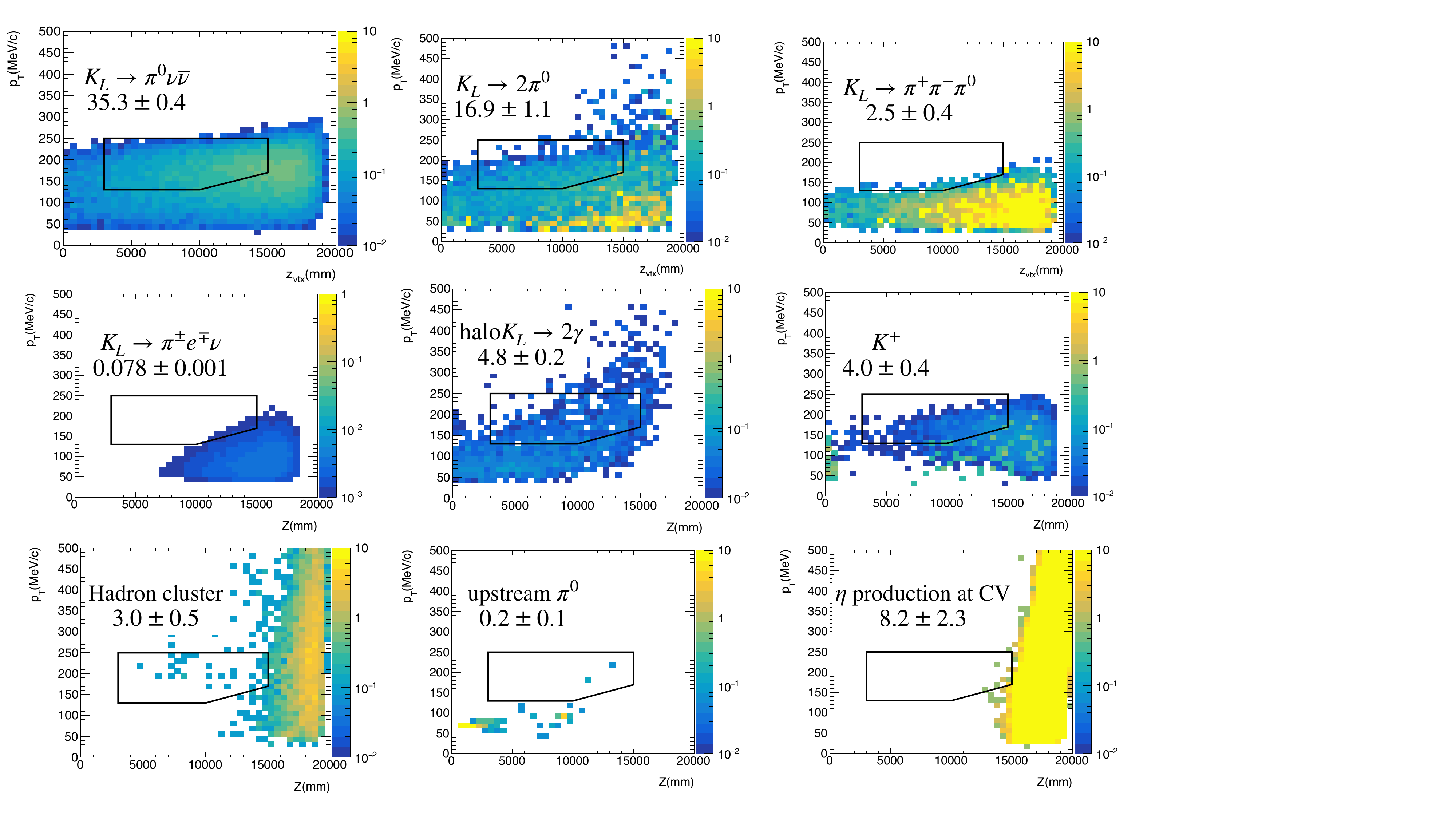}
\caption{KOTO-II distributions of the signal and backgrounds in the $z$--$p_T$ plane (simulation).\label{fig:ptz}}
\end{figure}

Including kinematic event selections,
the expected number of signal events is 35 if the SM value for the branching ratio is assumed, with a running time of $3\times 10^7~\mathrm{s}$
with a 100-kW beam incident in the T2 target.
The expected number of total background events is 40.
Distributions of the signal and background simulated events in the $z$--$p_T$ plane are shown in Fig.~\ref{fig:ptz}.
The signal can be observed with $5.6\sigma$ significance.
The branching ratio can therefore be measured with a statistical error of 25\%, resulting in a precision of the CKM parameter $\eta$ of 12\%.
Deviations of the branching ratio by 40\% from the SM value would indicate NP at 90\% CL.

The design of the KOTO-II detector is progressing towards a realistic geometry, see Fig.~\ref{fig:det3d}.
Based on the size and the weight from the realistic geometry,
and a radiation shielding simulation,
the surrounding detector area is under design. 
Prototyping of a modular barrel detector and of
a calorimeter with photon incident angle determination capabilities are on-going.
A shashlyk counter is considered for the outer region of the calorimeter.
A low-gain avalanche photodiode detector is being considered for the in-beam charged veto counter.
The option of a straw-tube tracker behind the charged veto detector to measure charged tracks is being evaluated.
The Collaboration plans to submit a proposal for KOTO~II in JFY 2024, in order to realise the KOTO~II experiment in the 2030s.

\subsection{Kaon physics from other experiments: LHCb and its Upgrade 2}


The LHCb experiment~\cite{LHCb:2008vvz} at the LHC is optimised primarily for the study of decays of the short-lived beauty and charm hadrons. In addition to its primary objectives, LHCb has proven to be suitable to investigate strange physics, despite the very low $\mathcal{O}$(100~MeV/$c$) transverse momentum ($p_T$) of the decay products of kaons and hyperons. In the past, the main bottleneck for strange physics at LHCb was the trigger system, which was selecting only events with $p_T > \mathcal{O}({\rm GeV})$ at the hardware level, resulting in a trigger efficiency of $\epsilon_{\rm trig} \sim 1\%$ in Run~1 (2010--2012). In Run~2 (2015--2018), a significantly modified software trigger enabled an improvement of the trigger efficiency, especially for channels with muons in the final state, by about an order of magnitude $\epsilon_{\rm trig} \sim 18\%$ with further improvements limited by the hardware trigger system. In Run~3, which started in 2022, the upgraded LHCb detector is equipped with an entirely software-based trigger system which will boost the sensitivity to kaon and hyperon decays with trigger efficiencies close to 100\%. The possibility of fully exploiting the data will therefore rely only on the capability of strange hadrons triggers to cope with the allowed rates, which for most of the channels mentioned here should be feasible.
The large improvement in trigger efficiency will enable LHCb to fully profit from the large data sets that will become available in the coming years. The data collected so far by LHCb in Run 1 and 2 correspond to 10~fb$^{-1}$. About 50~fb$^{-1}$ are expected to be collected after LHCb Run 3 and 4 and there is interest in continuing the experiment at high luminosity with a future Upgrade, possibly reaching 300~fb$^{-1}$~\cite{LHCb:2018roe} after Run 5 and 6. Furthermore, the huge strangeness production cross section at the LHC, two to three orders of magnitude larger than that of heavy flavours, makes strange-hadron physics an increasingly strong research line at LHCb~\cite{AlvesJunior:2018ldo}, with several results already published and more in the pipeline.
LHCb has published
the strongest bound on the branching fraction of $K^0_S \rightarrow \mu^+\mu^-$ decays~\cite{LHCb:2020ycd}, the first $4.1\sigma$ evidence for the rare $\Sigma^+ \rightarrow p \mu^+\mu^-$ decay~\cite{LHCb:2017rdd}, and the best upper limit on the branching fraction of $K^0_{S(L)} \rightarrow \mu^+\mu^-\mu^+\mu^-$ reaching $\mathcal{O}(10^{-12})$ level for the $K^0_S$ mode~\cite{LHCb:2022tpr}. 

Detailed sensitivity studies show that improvements of at least an order of magnitude are possible with LHCb Upgrade 2 data~\cite{AlvesJunior:2018ldo}.
Focusing first on rare decays, the LHCb experiment will be able to constrain the  $K_S \to \mu^+ \mu^-$ branching fraction down to about the SM level of $\sim 5\times 10^{-12}$. 
One of the most interesting decays in the short term will be the $K^0_S \rightarrow \pi^0\mu^+\mu^-$ decay. The form factor $a_S$, governing the $K^0_S \rightarrow \pi^0\mu^+\mu^-$ process can be extracted from a measurement of the $K^0_S \rightarrow \pi^0\mu^+\mu^-$ branching fraction. A precise measurement of $a_S$ is crucial for the prediction of its long-lived partner $K^0_L \rightarrow \pi^0\mu^+\mu^-$ decay, which is a very sensitive probe of physics beyond the SM.
The $K_S$ mode is currently known to only about $50\%$ precision from measurement by the NA48/1 collaboration, $\Br_\text{SM}(K^0_S \rightarrow \pi^0\mu^+\mu^-) = (2.9^{+1.5}_{-1.2} \pm 0.2) \times 10^{-9}$~\cite{NA481:2004nbc}. A more precise measurement of this branching fraction will result in an improved prediction of $K^0_L \rightarrow \pi^0\mu^+\mu^-$ and ultimately in improved BSM constraints that can be derived from it. The sensitivity of LHCb to $K^0_S \rightarrow \pi^0\mu^+\mu^-$ decays has been studied, demonstrating that significant improvements are possible depending on the trigger efficiency already with 10~fb$^{-1}$ of Upgrade data~\cite{Chobanova:2016laz}. This puts LHCb in a unique position to provide more information about this decay mode. The analysis of $K^0_S \rightarrow \pi^0\mu^+\mu^-$ decays can also be extended to other decays such as $K^0_S \rightarrow \gamma\mu^+\mu^-$, $K^0_S \rightarrow X\mu^+\mu^-$, $K^0_S \rightarrow X\pi\mu$, where $X$ is a scalar or vector particle. The search of lepton-flavour-violating $K_S \rightarrow \mu e$ decays can also be addressed by LHCb providing world-best limits for that mode.

A second group of decays which is gradually becoming more promising is the set of 4-body leptonic decays of the neutral kaon. No experimental constraints are present on the $K^0_S$ modes except for the recent limit on the $K^0_{S} \to \mu^+\mu^-\mu^+\mu^-$ mode provided by LHCb~\cite{LHCb:2022tpr}. Even though the rates for these decays are expected to be very low in the SM ($\Br(K^0_S \to e^+e^-e^+e^-)\sim 10^{-10}$, $\Br(K^0_S \to \mu^+\mu^-e^+e^-)\sim 10^{-11}$, $\Br(K^0_S \to \mu^+\mu^-\mu^+\mu^-)\sim 10^{-14}$), any sensitivity approaching the SM rates would be a test of NP, for example probing dark photons models. The prospects for such decays at LHCb will allow us to scan most of the allowed range in BSM models and get very close to the SM sensitivity if no signal is found.

Semileptonic hyperon decays can also be studied at LHCb. These decays profit from the relatively-high branching fractions, around $\Br\sim \mathcal{O}(10^{-4})$, which, coupled with the large strange hyperon production rates at the LHC, results in huge yields at LHCb. 
More comprehensive studies assessing the prospects for measurements with strange hadrons at LHCb can be found in Ref.~\cite{AlvesJunior:2018ldo}, using approximate simulations of the LHCb detector. A range of decays have been studied from $K^0_S$ to hyperons, showing that LHCb will be in a position to give significant contributions to strange-hadron physics in the near future.

\subsection{Discussion: Current and future experiments }

\subsubsection*{HIKE}
The measurement of $\Br(K^+ \rightarrow \pi^+ \nu\bar\nu)$ with a precision matching the core theoretical component of about 5\% (cf. Sec.~\ref{sec:gold-plated theory}) is uniquely interesting since it allows access to very high energy scales and can constrain or reveal several BSM models (see Section~\ref{sec:BSM}). 
Beyond the branching ratio measurement, the nature of the decay can be established by studying the kinematic distributions of signal candidates. 
Any BSM (scalar or tensor) contribution would not interfere with the SM contribution and therefore would manifest itself in an additive way. This means the measured kinematic distributions 
should be a sum of the SM plus BSM contributions. 
The investigation of the nature of the decay can probe fundamental properties of the SM. For example if the neutrinos are purely left-handed, as predicted by the SM, 
then the $K^{+}\rightarrow\pi^{+}\nu\bar{\nu}$ decay should be purely vector in nature, however if there is evidence of a different nature of the decay this indicates the presence of BSM,
almost certainly including lepton-number-violating operators (see Section~\ref{sec:bsm-dicuss}). 

Regarding other rare $K^+$ decays, the measurement of form factors in $K^{+}\rightarrow\pi^{+}\ell^{+}\ell^{-}$ decays is addressed also by theoreticians, with the foreseen lattice precision on form factors being about 10\% (cf. Sec.~\ref{sec:Lattice}) on the timescale of HIKE. The angular distribution analysis of these decays is also of theoretical interest, in relation to the corresponding one in the equivalent 
$B$-physics channel.
Generally, there is a definite theory interest in the differential studies of decay modes, for example the $m_{ee}$, $m_{\mu\mu}$ and $m_{\gamma\gamma}$ spectra (the latter from $K^+\to\pi^+\gamma\gamma$) since they, including cusp effects, could give access to BSM including exotica (see Section~\ref{sec:exotica}).

The study of the $K_L \to \pi^0\ell^+\ell^-$ decay is also important since it gives access to short-distance BSM effects in the photon coupling via the tau loop~\cite{isidori-seminar} that are not already included in $K \to\pi\nu\bar\nu$, see Sec.~\ref{sec:BSM_potential}.
Besides, the study of $K_{L}$ beta decays could allow a measurement of the $K^{+}$ mass, since the branching ratio is directly proportional to the fifth power of the mass difference $M_{K^0}-M_{K^+}$. This would help to resolve a long-running tension in the $K^{+}$ mass measurement. Sensitivity studies will be performed for HIKE Phase 2. Similar studies at KOTO~II are ongoing to see if some complementary investigations may be performed at J-PARC.

In HIKE, since both kaon and dump modes are foreseen, FIPs can be studied both in production and decay mode and could lead to independent self-contained identification of FIPs (without the need of further experiments).
The HIKE programme in particular offers an important opportunity to study FIP signatures since it has precision tracking and PID detectors. In contrast, if a FIP signature were discovered at a dedicated beam dump experiment, in principle a fixed-target experiment with tracking and PID would be needed to characterise it. 

Besides, in addition to the most promising channels, very rare decays such as $K^{+}\rightarrow\pi^{+}\ell^{+}\ell^{-}\gamma$ can be studied at HIKE, with very good possibilities to search for evidence of ALPs, with potentially different couplings. 
Other relevant channels to look for FIPs are $K_L \rightarrow 2\gamma, 4\gamma, 2 e, 4 e, \pi \pi e e, \pi \pi \gamma, 2\gamma 2e$ (see Section~\ref{sec:exotica}).

\begin{figure}[h]
\begin{center}
\includegraphics[width=0.495\textwidth]{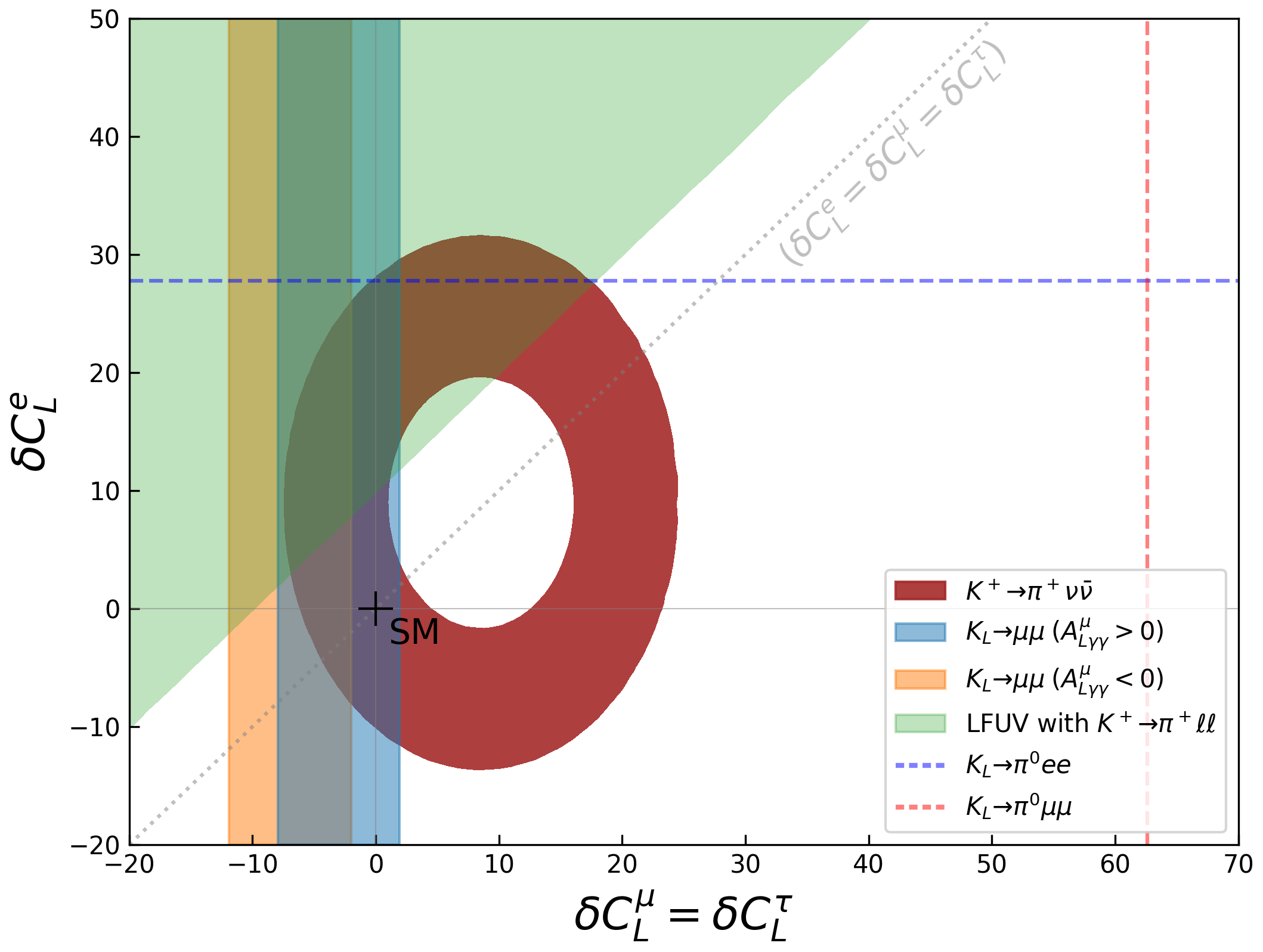}
\includegraphics[width=0.495\textwidth]
{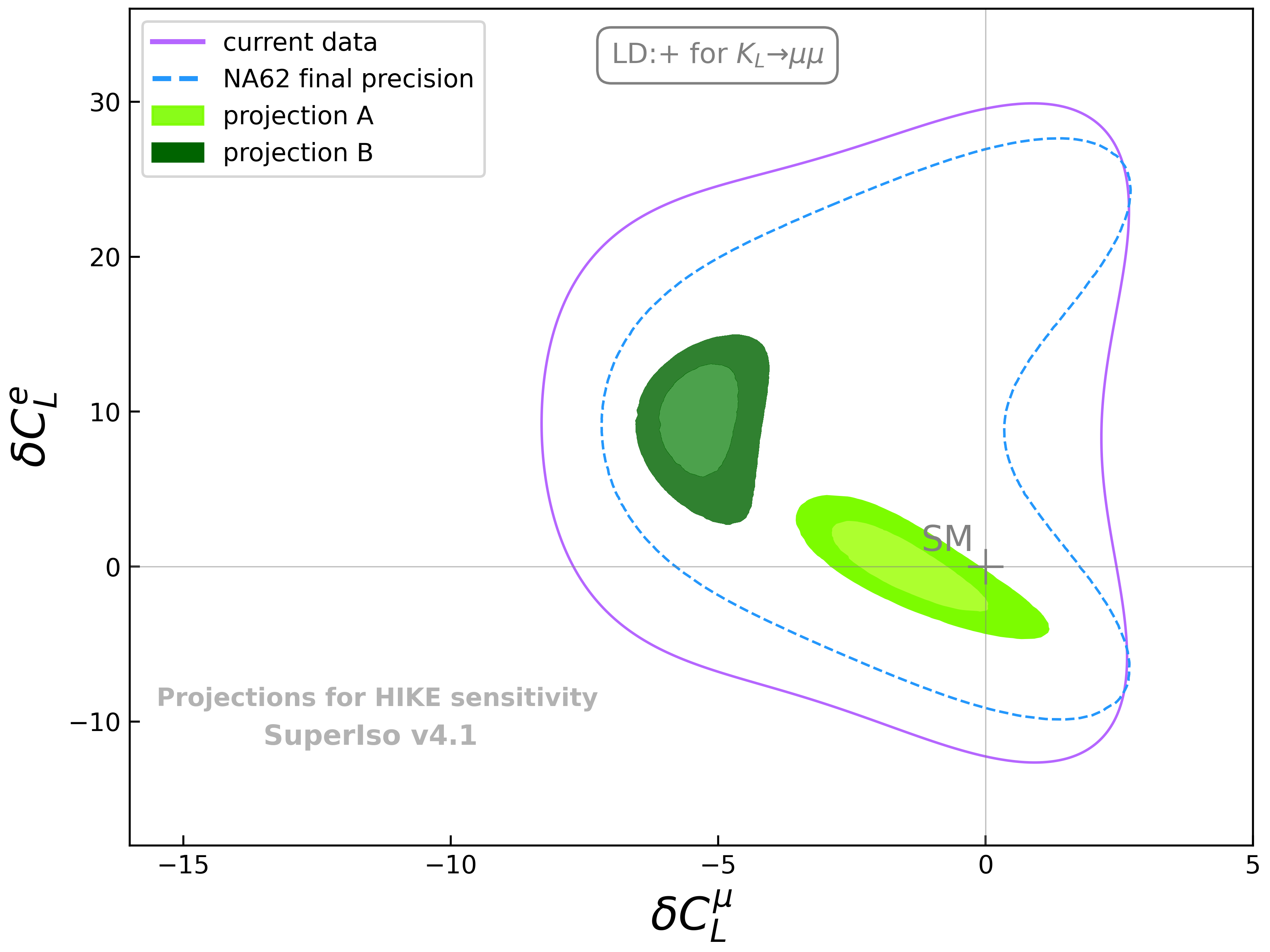}
\vspace{-3mm}
\caption{Bounds on LFU violating new physics contributions to Wilson coefficients from individual observables in the kaon sector (left). See Fig.~7 in Ref.~\cite{DAmbrosio:2022kvb} for further information.
Global fits in the Wilson coefficients plane with current data (purple contours) and the full projected scenarios (green regions) at the end of HIKE Phases~1 and~2 (for one choice of the two possible signs of the LD contributions to $K_L\to\mu^+\mu^-$). The blue dotted curve represents the NA62 projection at the end of 2025. For further details of the theory approach, and the full list of inputs, see ~Ref.~\cite{DAmbrosio:2022kvb,Neshatpour:2022fak,dambrosio2023}.
\label{fig:dambrosio}}
\end{center}
\end{figure}

The rare-kaon decays investigated by HIKE offer the possibility to search for BSM physics with a global-fit technique, for example, in the context of lepton flavour universality (LFU) tests~\cite{DAmbrosio:2022kvb}.
In the SM, the three lepton flavours ($e$, $\mu$ and $\tau$) have exactly the same gauge interactions and are distinguished only through their couplings to the Higgs field and hence the charged lepton masses. BSM models, on the other hand, do not necessarily conform to the lepton-flavour-universality hypothesis and may thereby induce subtle differences between the different generations that cannot be attributed to the different masses. Among the most sensitive probes of these differences are rare kaon decays with electrons, muons or neutrinos in the final state. For BSM scenarios with LFU violating effects, focusing on the case where the NP effects
for electrons are different from the those for muons and taus, bounds to Wilson coefficients from individual observables in the kaon sector are shown in Fig~\ref{fig:dambrosio} (left). A combined fit of all the decay modes is then performed~\cite{DAmbrosio:2022kvb, dambrosio2023}.
Projections based on the fits require assumptions for both the possible future measured (central) values as well as the experimental precision. For the latter, the expected long-term experimental precision is considered, while for the central values two scenarios are assumed: projection (A) where predicted central values for observables with only an upper bound available are taken to be the same as the SM prediction while for measured observables the current central values are used; projection (B) where the central values for all of the observables are projected with the best-fit points obtained from the fits with the existing data. The result of a combined fit of all the decay modes~\cite{Neshatpour:2022fak,dambrosio2023} is shown in Fig~\ref{fig:dambrosio} (right). It is evident that the combined measurements foreseen at HIKE, when taken together, have a larger potential to show a clear deviation from the SM or to strongly constrain the parameter space available to BSM physics than the single measurements taken in isolation.

\subsubsection*{KOTO and KOTO~II}
As mentioned in the KOTO and KOTO~II talks, the achievable sensitivities for the $K_L\to\pi^0\nu\bar\nu$ search will be better than $10^{-10}$ in KOTO (around $(5\text{--}8)\times 10^{-11}$, depending on the running time and the beam power) and $8\times10^{-13}$ with the running time of $3\times 10^7$~seconds and the beam power of 100~kW in KOTO-II.
Although KOTO and KOTO~II are designed and optimised for the flagship mode $K_L\to\pi^0\nu\bar\nu$, the physics programme can be extended to the following channels. It should be noted that the KOTO setup has no tracker and therefore the programme is basically focused on $K_L$ decays into photons. Another feature is the hermetic veto system, which enables one to search for the decay modes including invisible particles.
\begin{description}
\item[2 photons + invisible particle(s)]~\\
A natural extension in KOTO and KOTO~II is the $K_L\to\pi^0 X_{\rm inv}$ search, where $X_{\rm inv}$ represents any invisible particles. The sensitivity will be almost the same as that of the $K_L\to\pi^0\nu\bar\nu$ search. Given the reconstruction method and the expected sensitivity, the $K_L\to\pi^0 X_{\rm inv}$ search in KOTO~II would be limited by the backgrounds coming from $K_L\to\pi^0\nu\bar\nu$ decays.

\item[4 photons]~\\
In KOTO, the search for $K_L\to XX$, $X\to\gamma\gamma$ was performed using data taken in 2018~\cite{KOTO:2022lxx}. The upper limit, depending on the $X$ mass $(m_X)$, was set to be $(1\text{--}4)\times 10^{-7}$ for $40<m_X<110$~MeV/$c^2$ and $(1\text{--}2)\times 10^{-6}$ for $210<m_X<240$~MeV/$c^2$ at 90\% confidence level.

\item [4 photons + invisible particle(s)]~\\
A search for $K_L\to\pi^0\pi^0X_{\rm inv}$ was performed in the KEK E391a experiment, which is the predecessor of KOTO, conducted at the KEK 12~GeV proton synchrotron.
The upper limit for the branching ratio of $K_L\to\pi^0\pi^0\nu\bar\nu$ was set to be $8.1\times 10^{-7}$ at 90\% confidence level~\cite{E391a:2011aa}. The upper limit on $K_L\to\pi^0\pi^0X_{\rm inv}$ was also set varying from $7.0\times 10^{-7}$ to $4.0\times 10^{-5}$ for the mass of $X$ ranging from 50~MeV/c$^2$ to 200~MeV/c$^2$.
In order to improve this limit, a significant reduction of $K_L\to 3\pi^0$ backgrounds is needed.

\item[6 photons]~\\
$K_L\to\pi^0\pi^0 X$, $X\to\gamma\gamma$ corresponds to a peak search in the $m_{56}$ distribution, where $m_{56}$ represents the invariant mass of the photon pair other than $2\pi^0$. Events in the region other than the $\pi^0$ mass peak also come from $K_L\to 3\pi^0$ due to wrong combinations of photons in the event reconstruction. In KEK E391a, the search was done in a particular $X$ mass region, and the upper limit was set to be $(0.2\text{--}1)\times 10^{-6}$ at 90\% confidence level for $194<m_X<219$~MeV/$c^2$~\cite{E391a:2008grj}. A feasibility study in KOTO is under way.

\item[3 photons]~\\
KOTO performed the first search for the $K_L\to\pi^0\gamma$ decay, which is forbidden by Lorentz invariance, using data taken in 2016--2018. With a single event sensitivity of $7.1\times 10^{-8}$, the upper limit was set to be $1.7\times 10^{-7}$ at 90\% confidence level~\cite{KOTO:2020bhx}. The number of backgrounds was estimated to be 0.34, dominated by the $K_L\to 2\pi^0$ contribution, and thus further background reduction is needed to improve the limit.

\item[4 electromagnetic particles]~\\
Even though KOTO has neither a tracker nor a spectrometer, there is a possibility that a decay mode whose final state includes only electromagnetic particles can be reconstructed by using the energy and position information measured by the calorimeter. One of the interesting decay modes is $K_L\to\pi^0 e^+ e^-$. The vertex reconstruction can be done by the same method as used in the $K_L\to\pi^0\nu\bar\nu$ analysis with an assumption that two photons come from a $\pi^0$ decay, and then the invariant mass of two photons and $e^+$ and $e^-$ can be calculated.
KOTO is planning to collect a dataset for a feasibility study.
Additional constraints will be needed to reject the $K_L\to\gamma\gamma e^+e^-$ background.
In KOTO II, a feasibility study to install a tracker behind the charged-particle veto detector in front of the calorimeter is under discussion.

\end{description}


\subsubsection*{LHCb and its Upgrade 2} 

The LHCb experiment is currently unique in being sensitive to $K_S$ and hyperons~\cite{AlvesJunior:2018ldo}. In principle the experiment is sensitive also to $K^+$ and $K_L$ decays, however their long lifetime reduces their acceptance considerably by three orders of magnitude.

On hyperons, LHCb can probe $\Sigma^+$, $\Lambda$, $\Xi$ and $\Omega$ decays. LHCb will be able to measure not only the $\Sigma^+ \to p \mu^+ \mu^-$ integrated and differential branching fraction but also the CP violation asymmetry and the forward backward asymmetry. 
In a complementary way, the $\Sigma^+ \to p e^+ e^-$ decay will allow the study of $\Sigma^+ \to p \gamma$. 
In addition, all the other $s\to d \ell \ell$ transitions in the mentioned hyperons will be probed. 
Besides, $\Delta S=2$ decays (e.g., $\Xi \to p \pi$) can be probed and offer sensitivity to models not yet constrained by kaon mixing. 

In addition to rare decays, a programme of semileptonic measurements can be performed, starting from $K_S \to \pi\mu \nu$ and $\Lambda \to p \mu \nu$, 
where recent measurements from KLOE and BESIII have significantly improved the precision. 
Semileptonic hyperon decays can be studied in detail at LHCb and its Upgrades.
An improved measurement of these modes will be challenging due to the high levels of contamination from physics backgrounds but will offer high sensitivity to helicity-suppressed NP contributions~\cite{Chang:2014iba}. Despite the challenges, LHCb is expecting to achieve precision measurements of the branching fraction of $\Lambda \rightarrow p \mu^- \nu_\mu$, $\Xi^- \rightarrow \Lambda \mu^- \nu_\mu$, and $\Xi^- \rightarrow \Sigma^0 \mu^- \nu_\mu$ decays, as well as to perform sensitive searches for $\Xi^0 \rightarrow \pi^+\pi^-X$ and $\Xi^0 \rightarrow \mu^+\mu^-\pi^-X$ decays. 

The possibility of studying the interference of the $K_S$ and $K_L$ in  dimuon decays seems at the moment out of reach experimentally, but a future dedicated experiment could be thought about.

\subsection{Discussion: Complementarity between experiments}

HIKE and LHCb are complementary, with HIKE studying $K^{+}$ and $K_{L}$ and LHCb being primarily sensitive to $K_{S}$ and hyperons, and synergic in such that $K_S$ parameters can give input for theory calculations of key $K_L$ decays.

HIKE phase 2 and KOTO~II both study $K_{L}$ decays but with complementary physics goals: the former primarily investigating $K_L\to\pi^{0}\ell^{+}\ell^{-}$ decays using a detector with precise tracking information, and the latter highly optimised to study the $K_{L}\to\pi^{0}\nu\bar{\nu}$ decay mode. While KOTO~II is studying sensitivity to other $K_{L}$ modes, it is highly likely the specific design of the experiment and the crucial optimisation for the challenging decay $K_{L}\to\pi^{0}\nu\bar\nu$ will make it difficult to reach the precision HIKE Phase~2 can achieve. Nevertheless, in some modes some complementary and competitive measurements may be made to strengthen the cross validation of results in the community.
Indeed HIKE has a unique
advantage for channels with charged particles, in its powerful tracking and
PID system which are essential to reach the precision necessary for the observation of $K_L \rightarrow \pi^0 l^+ l^-$ including background suppression.

\subsection{Monte Carlo/QED contributions to the simulation and measurements of kaon physics}
\label{sec:MC}

When extracting information on strong or weak interaction dynamics from data, the particle spectra are affected by electromagnetic interactions among the particles involved in the studied process, as well as the real emission of additional photon quanta.
These distorted spectra can be corrected by means of the so-called radiative corrections.

The extra photons above the given experimental sensitivity threshold can be simulated in the Monte Carlo (MC), and the effects of the soft photons can be accounted for by using corrections to the spectrum of the non-radiative process.
All relevant contributions at the given order should be identified and calculated explicitly since approximate calculations tend to miss delicate cancellations among contributions.
The overall size of the QED effects in a given case can be estimated by integrating over the allowed energy range and emission angles of the additional photon(s), obtaining thus the (one-photon-)inclusive radiative corrections.
Needless to say, taking care of the radiative effects (tails) at the MC level leads to a desired agreement between data and MC and stable analyses with respect to the applied cuts.

Although the effect of the QED radiative corrections can seem negligible at first sight, being typically at the intuitive $\sim\!1\,\%$ level in the case of the integral decay width, this is no longer true for invariant-mass spectra or extracted hadronic parameters.
By studying one-photon-inclusive corrections to differential decay widths (or, in particular, Dalitz plots), one can easily encounter effects that are $\mathcal{O}(10\,\%)$.
Since the hadronic parameters like form-factor slopes can be comparable in size with the QED effects, corrections $\mathcal{O}(100\,\%)$ are not rare to occur.
The Dalitz decay of the neutral pion ($\pi^0\to e^+e^-\gamma$) is a prominent example.
One half of the slope of the next-to-leading-order (NLO) QED inclusive radiative corrections (given in terms of the normalized electron--positron invariant mass squared), which gives, in turn, an estimate on the size of the correction to the form-factor slope $a_\pi\simeq M_\pi^2/M_\rho^2\approx\!3\,\%$, amounts to about twice its size, approximately $-6\,\%$, and would need to be {\em subtracted} from the uncorrected measured value~\cite{Tupper:1983uw,Husek:2015sma}.
One finds corrections of similar size in processes like $\eta\to e^+e^-\gamma$~\cite{Husek:2017vmo} or $\Sigma^0\to\Lambda e^+e^-$~\cite{Husek:2019wmt}.

Naturally, the above considerations apply also to the kaon sector.
With increasing precision in the measurement of $K^+\to\pi^+\ell^+\ell^-$ form-factor parameters, lepton-flavour universality can be tested.
The QED contribution to the electron channel is expected to be rather large as compared to the muon channel, and ignoring these effects could lead to misinterpretation of the results.
Since the difference of the form-factor parameters obtained from respective channels directly relates to the associated LECs~\cite{Crivellin:2016vjc}, the QED part must be subtracted appropriately to obtain correct bounds.

The persisting tension in the $K^+\to\pi^0e^+\nu\gamma$ ($K_{e3\gamma}$) decay~\cite{NA62:2023lnp} may be also related to the underestimated size of the radiative corrections in the theoretical estimate of the inclusive ratio \mbox{$R=\Br(K_{e3\gamma(\gamma)}, E_\gamma^*>30\,\text{MeV}, \theta_{e\gamma}^*>20^\circ)/\Br(K_{e3(\gamma)})=0.640(8)\,\%$}~\cite{Fearing:1970iw,Bijnens:1992en,Kubis:2010mp} since there is no reason to expect that the intuitive $\sim\!1\,\%$ correction would apply for the limited phase space occurring in the numerator, the cuts there being designed to suppress the effects of the dominating inner-bremsstrahlung part.
The effects of radiative corrections are expected to be suppressed in the muon mode, which is related to the fact that the structure-dependent part becomes more important compared to the $K_{e3\gamma}$ case~\cite{Holstein:1990gu}.
It would be thus interesting to look at the `exclusive' measurement of the $K_{e3\gamma}$ decay or study the $K_{\mu3\gamma}$ mode.
On the experimental side, the MC simulation of the extra photons can be improved by providing the $K_{\ell3\gamma\gamma}$ generator.